\documentclass[useAMS,usenatbib]{mn2e}

\usepackage{color}
\usepackage{graphicx}
\usepackage{graphics}
\usepackage{rotating}
\usepackage{amsmath}        
\usepackage{amssymb}        
\usepackage{epsfig}        

\newcommand{\apjs}{ApJS}

\newcommand{\degree}{$^{\circ}$}

\newcommand{\perbeam}{\,beam$^{-1}$}
\newcommand{\perbm}{\,bm$^{-1}$}

\title[Disc-jet coupling in H1743-322]{Disc-jet coupling in the 2009 outburst of the black hole candidate H1743-322}

\author[J.~C.~A.~Miller-Jones et al.]
 {J.~C.~A.~Miller-Jones,$^{1,2}$\thanks{email: james.miller-jones@curtin.edu.au}
 G.~R.~Sivakoff,$^{3,4}$ D.~Altamirano,$^{5}$ M.~Coriat,$^{6,7}$ S.~Corbel,$^6$ \and V.~Dhawan,$^{8}$ H.~A.~Krimm,$^{9,10}$ R.~A.~Remillard,$^{11}$ M.~P.~Rupen,$^{8}$ D.~M.~Russell,$^{5}$ \and R.~P.~Fender,$^{7,5}$ S.~Heinz,$^{12}$ E.~G.~K\"ording,$^{13,6}$ D.~Maitra,$^{14}$ S.~Markoff,$^{5}$ S.~Migliari,$^{15}$ \and C.~L.~Sarazin,$^4$ and V.~Tudose$^{16}$ \\
$^1$International Centre for Radio Astronomy Research - Curtin University, GPO Box U1987, Perth, WA 6845, Australia\\
$^2$NRAO Headquarters, 520 Edgemont Road, Charlottesville,
 VA, 22903, USA\\
$^3$Department of Physics, University of Alberta, Room 238 CEB, Edmonton, AB T6G 2G7, Canada \\
$^4$Department of Astronomy, University of Virginia, P.O. Box 400325, Charlottesville, VA 22904, USA \\
$^5$Astronomical Institute `Anton Pannekoek', University of Amsterdam, P.O. Box 94249, 1090 GE Amsterdam, the Netherlands \\
$^{6}$Universit\'e Paris Diderot and Service d'Astrophysique, UMR AIM, CEA Saclay, F-91191 Gif-sur-Yvette, France \\
$^{7}$School of Physics and Astronomy, University of Southampton, Highfield SO17 IBJ, England\\
$^{8}$NRAO Domenici Science Operations Center, 1003 Lopezville Road, Socorro, NM 87801, USA \\
$^9$NASA/Goddard Space Flight Center, Greenbelt, MD 20771, USA\\ 
$^{10}$USRA, 10211 Wincopin Circle, Suite 500, Columbia, MD 21044, USA\\
$^{11}$MIT Kavli Institute for Astrophysics and Space Research, Building 37, 70 Vassar Street, Cambridge, MA 02139, USA \\
$^{12}$Astronomy Department, University of Wisconsin-Madison, 475. N. Charter St., Madison, WI 53706, USA \\
$^{13}$Department of Astrophysics, IMAPP, Radboud University Nijmegen,
Heyendaalseweg 135, 6525 AJ, Nijmegen, The Netherlands\\
$^{14}$Department of Astronomy, University of Michigan, Ann Arbor, MI 48109, USA \\
$^{15}$Departament d'Astronomia i Meteorologia, Institut de Ci\`ences del Cosmos (ICC), Universitat de Barcelona (IEEC-UB), \\Mart\'{\i} i Franqu\`es 1, E-08028 Barcelona, Spain\\
$^{16}$Netherlands Institute for Radio Astronomy, Oude Hoogeveensedijk 4, 7991 PD Dwingeloo, the Netherlands \\
}
\begin{document}

\date{Accepted 2011 December 2. Received 2011 November 30; in original form 2011 July 22}

\pagerange{\pageref{firstpage}--\pageref{lastpage}} \pubyear{2012}

\maketitle

\label{firstpage}

\begin{abstract}
We present an intensive radio and X-ray monitoring campaign on the 2009 outburst of the Galactic black hole candidate X-ray binary H1743-322.  With the high angular resolution of the Very Long Baseline Array, we resolve the jet ejection event and measure the proper motions of the jet ejecta relative to the position of the compact core jets detected at the beginning of the outburst.  This allows us to accurately couple the moment when the jet ejection event occurred with X-ray spectral and timing signatures.  We find that X-ray timing signatures are the best diagnostic of the jet ejection event in this outburst, which occurred as the X-ray variability began to decrease and the Type C quasi-periodic oscillations disappeared from the X-ray power density spectrum.  However, this sequence of events does not appear to be replicated in all black hole X-ray binary outbursts, even within an individual source.  In our observations of H1743-322, the ejection was contemporaneous with a quenching of the radio emission, prior to the start of the major radio flare.  This contradicts previous assumptions that the onset of the radio flare marks the moment of ejection.  The jet speed appears to vary between outbursts, with a possible positive correlation with outburst luminosity.  The compact core radio jet reactivated on transition to the hard intermediate state at the end of the outburst, and not when the source reached the low hard spectral state.  Comparison with the known near-infrared behaviour of the compact jets suggests a gradual evolution of the compact jet power over a few days near the beginning and end of an outburst.
\end{abstract}


\section{Introduction}

The nature of the coupling between inflow and outflow in accreting compact objects has been the focus of intensive study in recent years.  Galactic X-ray binary sources provide ideal laboratories in which to probe this disc-jet coupling, since they pass through a number of well-defined accretion regimes as they evolve through their full duty cycles on timescales of months to years.  Thus we can study a range of distinct accretion states in the same object, hence controlling for the effects of compact object mass, donor type and environment on the formation of jets in different accretion regimes.

The properties of the accretion flow are best studied in the X-ray band, and have been categorised into a number of canonical states \citep[see, e.g.][for a review]{Bel10}, which are identified by specific X-ray spectral and timing characteristics.  Black hole X-ray binaries spend the majority of their time in a very low luminosity ($<10^{33.5}$\,erg\,s$^{-1}$) quiescent state \citep{McC06}.  At the beginning and end of an outburst, such systems are observed in the low/hard state (LHS), with an X-ray spectrum dominated by a fairly hard power law component (photon index $\Gamma\sim1.5$), and a power density spectrum showing strong band-limited noise with integrated fractional rms variability $\sim0.3$.  During the peak phase of an outburst, a black hole X-ray binary will make a transition through a set of intermediate states to the high/soft state (HSS).  The HSS has an X-ray spectrum dominated by soft thermal blackbody emission from a geometrically thin, optically thick accretion disc, together with a weak, steep ($\Gamma\gtrsim2.5$) power-law component, and showing a power density spectrum with a much reduced fractional rms variability of $\lesssim0.1$.  The intermediate states were classified by \citet{Hom05} into two distinct classes, the hard and soft intermediate states (HIMS and SIMS respectively), distinguished by both the photon index of the power law component and by their X-ray variability properties, notably a rapid decrease in the fractional rms variability and a transition between two types of quasi-periodic oscillations (QPOs) on moving from the HIMS to the SIMS.  At the end of the outburst, the X-ray luminosity decreases, and the source makes a reverse transition from the HSS through the HIMS back to the LHS.  The reduced luminosity of the soft-to-hard transition as compared to the hard-to-soft transition leads to an observed hysteresis in a hardness-intensity diagram (HID).  While there are alternative sets of state definitions, most notably those of \citet{McC06}, we have chosen to use the terminology outlined above primarily for consistency with much of the published literature on the coupling of radio and X-ray emission from black hole X-ray binaries, specifically the phenomenological picture presented by \citet*{Fen04} and \citet*{Fen09}.

The relativistic jets in black hole X-ray binaries are seen in two distinct forms.  Steady, compact jets have been directly imaged in hard X-ray spectral states \citep*[e.g.][]{Dha00,Sti01}, and are observed to have a flat or slightly inverted radio spectrum \citep{Fen01}.  They are best modelled as partially self-absorbed conical outflows \citep[e.g.][]{Bla79}, and are believed to carry away a significant, if not dominant, fraction of the power liberated by the accretion flow \citep*{Fen03,Gal05}.  The second, more spectacular, type of jet morphology is seen during outbursts, as bright, steep-spectrum, discrete, relativistically-moving knots of radio emission.  In most systems, these are only observed on milliarcsecond scales using very long baseline interferometry \citep[VLBI; e.g.][]{Tin95,Hje95,Han01}, although for the brightest systems, they may be observed on larger angular scales with connected-element interferometers such as the Very Large Array \citep*[VLA; e.g.][]{Mir94,Mar01}.  Optically thin radio emission may also be observed some distance downstream from the X-ray binary itself if the jets run into a particularly dense patch of the surrounding interstellar medium, causing particle acceleration at the sites of the resulting external shocks \citep[e.g.][]{Cor02,Cor05}.  Since this is dependent on the properties of the surrounding medium rather than the launching mechanism of the jets themselves, we will not focus on it in any detail.

The relationship between the radio emission (assumed to arise from the jets) and the X-ray emission from the accretion flow was considered in detail by \citet{Cor04}, who used the Australia Telescope Compact Array (ATCA) to sample the radio emission throughout the 2001 outburst of the black hole X-ray binary XTE J1650-500.  Comparing their data to similar outbursts in three other systems, they found that the hard state (corresponding to the LHS in our terminology) is associated with emission from a partially self-absorbed compact jet, which persists into the intermediate state (equivalent to the HIMS).  Ejection events are associated with transitions from the intermediate to the steep power law state (HIMS to SIMS transition), and the jet is suppressed in the thermal dominant state (HSS), although residual radio emission may be observed as a result of interactions between the ejecta and the surrounding environment.

 \citet{Fen04a} also made a detailed study of the relationship between the X-ray and radio emission (the disc-jet coupling) in a black hole X-ray binary, in this case focussing on the system GRS\,1915+105.  Although the X-ray spectral states in GRS\,1915+105 are non-standard, multi-wavelength observations by \citet{Mir98} also provided evidence that jet ejection events are triggered at the end of a phase of X-ray spectral softening.  This was confirmed by high-resolution observations of a large radio flare \citep{Fen99}, which measured the proper motions of the relativistic jets, showing them to be ejected as the source was moving from a relatively hard `plateau' state to a softer X-ray state.

\citet{Fen04} generalised this picture by synthesizing the existing data on the disc-jet coupling in several different black hole X-ray binaries into a `unified model'.  In its simplest form, this model holds that the steady, compact jets exist during the quiescent state and LHS, with a relatively low bulk velocity (Lorentz factor $\Gamma\lesssim 2$).  As the X-ray luminosity ($L_{\rm X}$) rises at the beginning of an outburst, so too does the jet power and hence the radio luminosity ($L_{\rm r}$), following the non-linear correlation $L_{\rm r}\propto L_{\rm X}^{0.7}$ \citep*{Cor03,Gal03}.  As the source moves into the intermediate states, the X-ray spectrum softens and the jet Lorentz factor increases, with the jet becoming unstable.  At a certain point in the HID, dubbed the `jet-line', the core jet switches off, and a radio flare may occur as the faster moving material ejected more recently catches up with the pre-existing, slower-moving flow, giving rise to internal shocks, which accelerate particles to produce the observed bright, optically thin, relativistically-moving ejecta.  Alternatively, the shocks could be caused by increased outflow flux at constant velocity, rather than by a change in the bulk velocity.  Supporting such a scenario, the disappearance of the non-thermal power-law component of the X-ray spectrum at this time has led to suggestions that the shocks could be caused by the ejection of the corona and its subsequent interaction with material from the pre-existing steady jets \citep*{Rod03,Vad03,Rod08}.  If the X-ray behaviour oscillates back and forth across the jet line, multiple radio flares may be observed, with the compact core jet reforming each time the spectrum becomes harder, and another ejection event occurring at the subsequent softening event.  During the HSS, the compact core jet is suppressed by over two orders of magnitude \citep{Cor11,Rus11}, and reforms in the final stages of the outburst as the source moves from a HIMS back to the LHS.

While many authors have used this model as a benchmark against which to compare their observational data \citep[e.g.][]{Cor04,Mil05,Fen06}, it was evaluated in detail by \citet{Fen09}, who attempted to couple the moment of jet ejection with particular X-ray spectral and timing features.  While they verified the main features of the previous picture, they were unable to identify any unique X-ray signature of the jet ejection event (the crossing of the jet line), probably due to the sparse radio coverage for most of the studied events.  These authors identified a need for more intensive radio coverage during the transitional phases of the outburst, particularly in the form of high angular resolution VLBI monitoring.  Such resolved radio imaging is crucial in accurately interpreting the jet emission, since the peak in the radio light curve may not correspond to the exact time when the moving knots were ejected from the core.  A clear example of this was identified by \citet{Ver87}, who found that the jet emission in SS\,433 reached a maximum when crossing a `brightening zone' located $\sim250$\,au downstream from the core.  Absorption effects, recollimation shocks, or the time delay required to form internal shocks could all prevent an accurate assessment of the disc-jet coupling when relying on radio flux density monitoring alone.

\subsection{H1743-322}
\label{sec:h1743}

First detected during an outburst in 1977 \citep{Kal77}, H1743-322 was rediscovered as a new source in 2003 by INTEGRAL \citep{Rev03}, only subsequently being associated with the previously known source \citep{Mar03}.  Since that time, seven smaller outbursts of the source have been observed \citep{Swa04,Rup05,Kal08,Kuu08,Kri09,Nak10,Kuu11}, at least one of which failed to make a full transition through all the canonical states \citep{Cap09}.

From its similarities with the dynamically confirmed black hole X-ray binary XTE J1550-564, the source was classified as a black hole candidate \citep{McC09}, although no mass function has yet been reported for the system.  From its location only 3.2\degree\ from the Galactic Centre, the source distance was assumed to be $\sim 8$\,kpc.  This was recently confirmed by \citet*{Ste12}, who derived a distance of $8.5\pm0.8$\,kpc by modelling the trajectories of the radio jets during the 2003 outburst.  \citet{McC09} measured quiescent magnitudes for the system of 17.1 at K-band and $>24$ at $i^{\prime}$ band, but with no firm distance determination and uncertain contributions from the jet and outer disc, the spectral type of the secondary is also unknown.  However, from the relation between orbital period and quiescent X-ray luminosity \citep{Gar01}, and their measured quiescent luminosity of $\sim3\times10^{32}$\,erg\,s$^{-1}$ (0.5--10\,keV), \citet{Jon10} inferred an orbital period of $\gtrsim10$\,h.

From the X-ray dipping behaviour observed by \citet{Hom05b} and \citet{Mil06}, the inclination angle of the disc in H1743-322 is believed to be relatively high, on the order of $\sim80$\degree\ to our line of sight.  \citet{Ste12} also determined the inclination angle of the radio jets to be $75\pm3$\degree, implying that the disc and jet axes are therefore relatively well aligned.  Narrow, variable and blueshifted absorption lines from highly-ionized Fe suggest a disk wind in this system \citep{Mil06}, which was subsequently found to be partially state-dependent, with the wind being stronger in the high/soft state \citep{Blu10}.  The mass outflow rate in the wind was estimated as being on the order of $3\times10^{17}$\,g\,s$^{-1}$, assuming a wind velocity of 300\,km\,s$^{-1}$.

The X-ray and radio behaviour of the source during the 2003 outburst were studied in detail by \citet{McC09}.  The event was complicated, with multiple transitions between the different X-ray spectral states, and several different radio flaring events.  Owing to the rapid X-ray variability, complicated outburst behaviour and sparse radio sampling, no detailed comparison of the jet-disc coupling with the phenomenological model of \citet{Fen04} was possible.

Several months after the 2003 outburst, moving radio emission was detected at arcsecond scales on either side of the central source, subsequently discovered to be associated with X-ray sources \citep{Cor05}.  These were interpreted as the interaction of the ejecta from the original outburst with the surrounding environment.  While the jets were found to have decelerated since the 2003 event, the observations placed constraints on the position angle of the jets and on their apparent velocity.

In this paper, we present detailed radio observations of the 2009 outburst of H1743-322.  The X-ray evolution of the outburst has already been analysed in detail by \citet*{Mot10} and \citet{Che10}, and the source was found to follow the canonical path through the HID, with a relatively simple evolution as compared to the 2003 event.  The dense radio coverage at high angular resolution, coupled with the simple, canonical behaviour in the X-ray band, allowed us to make a detailed study of the disc-jet coupling in this system, testing the model of \citet{Fen04} and addressing some of the issues raised by \citet{Fen09}.

\section{Observations and data reduction}
The outburst was monitored with the Proportional Counter Array (PCA) on board the {\it Rossi X-ray Timing Explorer} ({\it RXTE}), and its X-ray spectral and timing characteristics have been used by \citet{Mot10} to distinguish the various characteristic accretion states.  We monitored the outburst in the radio band using the VLA and Very Long Baseline Array (VLBA) under the auspices of the Jet Acceleration and Collimation Probe of Transient X-ray binaries (JACPOT XRB) program \citep{Mil11}, tracking the evolution of the radio emission through the entire outburst.  We also obtained data from the ATCA to improve our temporal coverage in the radio band.

\subsection{VLA}

After {\it Swift} reported a re-brightening of H\,1743-322 in the 15--50\,keV band from 2009 May 22 \citep{Kri09}, we made preliminary radio observations with the VLA, detecting the source at a level of 2.2\,mJy at 8.4\,GHz \citep{Mil09}.  On the basis of this detection, we triggered a long-term VLA monitoring campaign (program code AM991), tracking the radio emission from the source over 24 epochs as it made the transition from the LHS to the HSS, and again as it made the reverse transition back to the LHS (see Table~\ref{tab:vla} for more details).  The array was in its CnB and C configurations during the observations, with the configuration change occurring on 2009 June 16.  The monitoring was carried out at frequencies of 1.4, 4.9 and 8.4\,GHz, with two epochs of observation at 22.4\,GHz during the peak of the radio outburst in an attempt to better constrain the source spectrum when the source was brightest.  We observed in dual-polarization mode with 100\,MHz of bandwidth per polarization, split equally between two intermediate frequency (IF) pairs.  We used 3C\,286 as a primary calibrator, setting the amplitude scale according to the coefficients derived at the VLA in 1999 by staff at the National Radio Astronomy Observatory, as implemented in the 31Dec08 version of the Astronomical Image Processing System \citep[{\sc aips};][]{Gre03}.  The secondary calibrator was J1744-3116, at an angular separation of 1\degree\ from H1743-322.  Observations were carried out in fast-switching mode with a 4-min cycle time (1\,min on the calibrator, 3\,min on the target source) to minimize down time while maximizing the accuracy of the phase transfer at the low elevation of the sources.  Data reduction was carried out according to standard procedures within {\sc aips}.  The observations were typically of duration 1\,h split across all observing frequencies, and were simultaneous with monitoring at the VLBA wherever possible.  On a few occasions, the weather was sufficiently bad that the calibrator phase changed by several tens of degrees over the course of a 60-s scan.  At these epochs, the interpolation of phases to the target source was poor, and since H1743-322 was not bright enough for self-calibration we were unable to correct for this effect, leading to larger than expected errors on the measured source brightnesses.  In one instance, on 2009 July 4, the weather was sufficiently bad (phase changing by $>90$\degree\ within 60\,s) that the target source could not be detected.

\begin{table*}
\begin{center}
\small
\begin{tabular}{lclcccccc}
\hline\hline
Date & MJD$^a$ & X-ray$^b$ & $S_{\rm 1.4 GHz}$ & $S_{\rm 4.9 GHz}$ & $S_{\rm 8.4 GHz}$ & $S_{\rm 22.4 GHz}$ & $\alpha_{8.4,4.9}$ &  $\alpha_{8.4,1.4}$\\
& (d) & state & (mJy\perbm) & (mJy\perbm) & (mJy\perbm) & (mJy\perbm) & &\\
\hline
2009 May 27 & 54978.4 & $-$ & $-$ & $-$ & $2.24\pm0.05$ & $-$ & $-$ & $-$ \\
2009 May 30 & 54981.4 & HIMS & $2.03\pm0.30$ & $2.76\pm0.10$ & $2.73\pm0.01$ & $-$ & $-0.02\pm0.09$ & $0.16\pm0.08$\\
2009 Jun 2 & 54984.3 & HIMS & $1.80\pm0.30$ & $2.50\pm0.06$ & $2.48\pm0.04$ & $-$ & $-0.01\pm0.05$ & $0.18\pm0.09$\\
2009 Jun 7 & 54989.2 & HSS & $-$ & $5.7\pm0.4$ & $12.8\pm0.3$ & $-$ & $1.43\pm0.13$ & $-$\\
2009 Jun 9 & 54991.4 & SIMS to HSS & $-$ & $-$ & $2.20\pm0.80$ & $<1.41^c$ & $-$ & $-$\\
2009 Jun 11 & 54993.3 & HSS & $1.71\pm0.30$ & $0.90\pm0.08$ & $0.65\pm0.04$ & $<1.11^c$ & $-0.58\pm0.20$ & $-0.53\pm0.10$\\
2009 Jun 14 & 54996.2 & HSS & $<1.95^c$ & $0.43\pm0.10$ & $0.18\pm0.06$ & $-$ & $-1.56\pm0.78$ & $-$\\
2009 Jun 15 & 54997.3 & HSS & $-$ & $0.35\pm0.04$ & $0.38\pm0.06$ & $-$ & $-0.18\pm0.35$ & $-$\\
2009 Jun 16 & 54998.3 & HSS & $-$ & $-$ & $0.31\pm0.05$ & $-$ & $-$ & $-$ \\
2009 Jun 17 & 54999.2 & HSS & $-$ & $-$ & $0.16\pm0.05$ & $-$ & $-$ & $-$ \\
2009 Jun 18 & 55000.3 & HSS & $-$ & $-$ & $0.17\pm0.06$ & $-$ & $-$ & $-$ \\
2009 Jun 19 & 55001.3 & HSS & $-$ & $-$ & $0.19\pm0.06$ & $-$ & $-$ & $-$ \\
2009 Jun 23 & 55005.3 & HSS & $-$ & $-$ & $0.34\pm0.07$ & $-$ & $-$ & $-$ \\
2009 Jun 25 & 55007.2 & HSS & $-$ & $-$ & $0.35\pm0.06$ & $-$ & $-$ & $-$ \\
2009 Jun 27 & 55009.3 & HSS & $-$ & $-$ & $0.33\pm0.06$ & $-$ & $-$ & $-$ \\
2009 Jun 29 & 55011.2 & HSS & $-$ & $-$ & $0.24\pm0.06$ & $-$ & $-$ & $-$ \\
2009 Jun 30 & 55012.2 & HSS & $-$ & $-$ & $0.27\pm0.06$ & $-$ & $-$ & $-$ \\
2009 Jul 7 & 55019.2 & HIMS & $-$ & $-$ & $0.59\pm0.06$ & $-$ & $-$ & $-$ \\
2009 Jul 8 & 55020.3 & HIMS & $0.95\pm0.22$ & $0.65\pm0.08$ & $0.55\pm0.06$ & $-$ & $-0.32\pm0.30$ & $-0.31\pm0.14$ \\
2009 Jul 9 & 55021.3 & HIMS & $0.99\pm0.26$ & $0.48\pm0.09$ & $0.41\pm0.07$ & $-$ & $-0.27\pm0.47$ & $-0.49\pm0.18$\\
2009 Jul 12 & 55024.3 & HIMS & $1.30\pm0.27$ & $0.48\pm0.09$ & $0.34\pm0.06$ & $-$ & $-0.66\pm0.50$ & $-0.75\pm0.16$\\
2009 Jul 14 & 55026.2 & HIMS to LHS & $0.71\pm0.25$ & $0.62\pm0.08$ & $0.59\pm0.07$ & $-$ & $-0.09\pm0.32$ & $-0.11\pm0.20$ \\
2009 Jul 20 & 55032.2 & LHS & $1.25\pm0.27$ & $0.47\pm0.08$ & $0.63\pm0.05$ & $-$ & $0.52\pm0.35$ & $-0.38\pm0.13$\\
2009 Aug 6 & 55049.2 & LHS & $-$ & $<0.20^c$ & $0.19\pm0.05$ & $-$ & $-$ & $-$ \\
\hline
\end{tabular}
\end{center}
{\caption{\label{tab:vla}Log of the VLA observations, showing for each epoch the X-ray state of the source, the brightness of H1743-322 at each frequency, and the derived radio spectral index $\alpha$ where appropriate.
\newline \newline
$^a$ The reported MJD corresponds to the midpoint of the observations.
\newline
$^b$ X-ray states are those derived by \citet{Mot10}.  Where the radio data were taken between two X-ray observations in different states, we report the relevant transition.
\newline
$^c$ In the case of non-detections, we quote the $3\sigma$ upper limit to the source brightness.
}}
\end{table*}

\subsection{VLBA}

On detection of a radio source with the VLA, we triggered a campaign of observations with the VLBA under program code BM308, monitoring the source over ten epochs throughout its transition from the LHS to the HSS.  Once our VLA monitoring showed that the radio emission had faded below detectable levels for the VLBA, the high-resolution monitoring ceased until X-ray monitoring showed the onset of the reverse transition back to the LHS, whereupon we made a total of six further observations.

We observed in dual polarization mode at a primary observing frequency of 8.4\,GHz, although we also observed at 15\,GHz for the four epochs during the peak of the outburst when the radio emission was brightest.  Each epoch was of duration 5\,h, the maximum time for which the source was above the elevation limit of the antennas.  For all epochs except the last one, our recording rate was 256\,Mbps, corresponding to a total bandwidth of 32\,MHz per polarization.  For the final epoch, when our VLA monitoring showed that the source was well into the decaying LHS and very faint, we doubled the recording rate and observing bandwidth.  In the last ten epochs of observation, we spent 30\,min at the beginning and end of the observing run taking data on bright extragalactic calibrators distributed over a wide range of elevations across the sky (i.e.\ geodetic blocks).  This enabled us to better account for unmodeled clock and tropospheric phase errors using the {\sc aips} task DELZN, thereby improving the success of the phase transfer.

Owing to its location only 3.2\degree\ away from the Galactic Centre, the angular broadening caused by the strong scattering along the line of sight to the source creates a dearth of good, close VLBI calibrators, particularly at lower radio frequencies.  Based on experience from previous observing campaigns on H1743-322 in 2003, 2004 and 2008, we used the sources J1748-2907 and J1740-2929 as our primary phase reference sources, both believed to be extragalactic on account of their brightness temperatures, morphologies and steady fluxes \citep*{Bow01}.  For J1748-2907, we used the updated position of \citet{Rei04}.  Our phase referencing cycle time was 3 minutes, spending 2 minutes on H1743-322 and 1 minute on the calibrator source during each cycle.  Every seventh cycle, we substituted a scan on the target source for a scan on the check source J1744-3116 from the third extension to the VLBA Calibrator Survey \citep[VCS-3;][]{Pet05}, and then observed the fringe finder source J1733-1304.  Our assumed calibrator positions are given in Table~\ref{tab:calibrators}.  Despite the greater calibrator throw to J1748-2907 and J1740-2929, they were preferred to J1744-3116 as phase reference sources since the latter source is known to be highly scattered at 8.4\,GHz, reducing the number of antennas for which phase solutions could be derived.  

\begin{table}
\begin{center}
\small
\begin{tabular}{lccc}
\hline\hline
Calibrator & Right Ascension & Declination & Calibrator\\
& & & throw (\degree)\\
\hline
J1748-2907 & 17$^{\rm h}$48$^{\rm m}$45\fs6860 & $-29$\degr07\arcmin39\farcs404 & 3.2\\
J1740-2929 & 17$^{\rm h}$40$^{\rm m}$54\fs5275 & $-29$\degr29\arcmin50\farcs327 & 3.0\\
J1744-3116 & 17$^{\rm h}$44$^{\rm m}$23\fs5784 & $-31$\degr16\arcmin36\farcs2913 & 1.0\\
\hline
\end{tabular}
\end{center}
{\caption{\label{tab:calibrators}Assumed positions of the phase reference and check sources (all in J\,2000 coordinates).  The last column shows the angular separation between the calibrator and H1743-322.}}
\end{table}

\begin{table*}
\begin{center}
\small
\begin{tabular}{lclccllc}
\hline\hline
Date & MJD$^a$ & X-ray$^b$ & Frequency & Bandwidth & Phase reference & Geodetic & H1743-322 flux\\
& (d) & state & (GHz) & (MHz) & calibrator & block? & density$^c$ (mJy)\\
\hline
2009 May 28 & $54979.35\pm0.10$ & $-$ & 8.4 & 32 & J1748-2907, J1740-2929 & N & $2.0\pm0.2$\\
2009 May 30 & $54981.35\pm0.10$ & HIMS & 8.4 & 32 & J1748-2907, J1740-2929 & N & $2.1\pm0.3$\\
2009 Jun 2 & $54984.35\pm0.10$ & HIMS & 8.4 & 32 & J1748-2907, J1740-2929 & N & $1.8\pm0.3$\\
2009 Jun 5 & $54987.33\pm0.10$ & SIMS & 8.4 & 32 & J1748-2907, J1740-2929 & N & $5.7\pm0.3$\\
 & $54987.33\pm0.10$ & SIMS & 15.0 & 32 & J1748-2907, J1740-2929 & N & $2.2\pm0.4$\\
2009 Jun 6$^d$ & $54988.33\pm0.10$ & HSS & 8.4 & 32 & J1740-2929 & N & $6.4\pm0.3$\\
 & & HSS & & & & & $7.1\pm0.3$\\
 & $54988.34\pm0.09$ & HSS & 15.0 & 32 & J1748-2907 & N & $2.9\pm0.5$\\
2009 Jun 8 & $54990.33\pm0.10$ & SIMS & 8.4 & 32 & J1740-2929 & N & $<1.0$\\
 & $54990.34\pm0.09$ & SIMS & 15.0 & 32 & J1740-2929 & N & $<1.7$\\
2009 Jun 9 & $54991.33\pm0.08$ & HSS & 8.4 & 32 & J1740-2929 & Y & $<1.1$\\
 & $54991.33\pm0.09$ & SIMS to HSS & 15.0 & 32 & J1740-2929 & Y & $<1.1$\\
2009 Jun 11 & $54993.33\pm0.08$ & HSS & 8.4 & 32 & J1740-2929 & Y & $<0.8$\\
2009 Jun 13 & $54995.33\pm0.08$ & HSS & 8.4 & 32 & J1740-2929 & Y & $<1.0$\\
2009 Jun 15 & $54997.31\pm0.08$ & HSS & 8.4 & 32 & J1740-2929 & Y & $<0.8$\\
2009 Jul 9 & $55021.23\pm0.09$ & HIMS & 8.4 & 32 & J1748-2907 & Y & $<0.8$\\
2009 Jul 11 & $55023.25\pm0.08$ & HIMS & 8.4 & 32 & J1748-2907 & Y & $<0.8$\\
2009 Jul 12 & $55024.25\pm0.08$ & HIMS & 8.4 & 32 & J1748-2907 & Y & $<0.8$\\
2009 Jul 18 & $55030.23\pm0.08$ & LHS & 8.4 & 32 & J1748-2907 & Y & $<0.8$\\
2009 Jul 20 & $55032.06\pm0.06$ & LHS & 8.4 & 32 & J1748-2907 & Y & $<0.9$\\
2009 Aug 6 & $55049.18\pm0.09$ & LHS & 8.4 & 64 & J1748-2907 & Y & $<0.4$\\
\hline
\end{tabular}
\end{center}
{\caption{\label{tab:vlba}Log of the VLBA observations, showing for each epoch the X-ray state of the source, the observing frequencies and bandwidths, phase calibrators, and whether a geodetic block was used.
\newline \newline
$^a$ The reported MJD corresponds to the midpoint of the observations.  Error bars denote the duration of the observations either side of the midpoint, such that the total duration corresponds to twice the reported time uncertainty.
\newline
$^b$ X-ray states are those derived by \citet{Mot10}.
\newline
$^c$ Where the source was not detected, $5\sigma$ upper limits are given in all cases.  This shows the confidence with which we can rule out the detection of jet ejecta at an unknown position.  However, detection of the core can be ruled out at the $3\sigma$ level, i.e.\ 40 per cent lower than the tabulated figures, since the core position is known.
\newline
$^d$ At this epoch, 2 components were detected at 8.4\,GHz, on either side of the core position (see Fig.~\ref{fig:h1743_8ghz}).  Flux densities for both components are reported.  At 15.0\,GHz, only one component was detected, whose position was consistent with the western component in the 8.4\,GHz image.}}
\end{table*}

\begin{figure*}
\centering
\includegraphics[width=\textwidth]{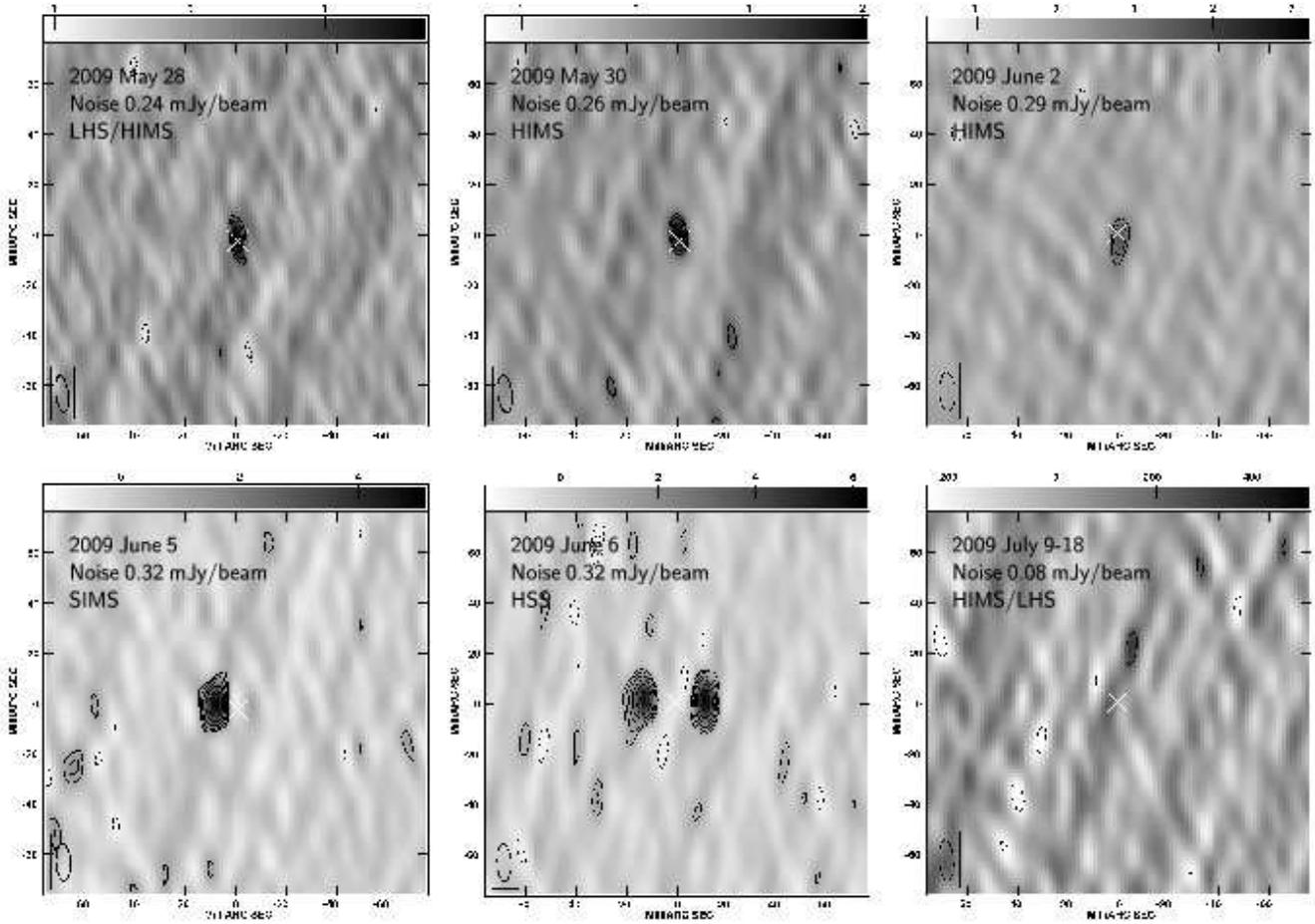}
\caption{8.4-GHz VLBA images of H1743-322.  Contour levels are at $\pm(\sqrt{2})^n$ times the rms noise level indicated in the top left corner of each image, where $n=-3,3,4,5...$  The grey scale shows the image flux density in units of mJy\perbeam, except for the stacked image of 2009 July 9--18, for which the units are $\mu$Jy\perbeam.  The white cross marks the weighted mean core position derived in Section~\ref{sec:core}.  We see an unresolved core jet in the first three epochs, followed by a two-sided ejection event.  The positional offset of the possible northwestern source in the image of 2009 July 9--18, together with the marginal nature of the detection, implies that we cannot convincingly associate this source with the compact core jet.}
\label{fig:h1743_8ghz}
\end{figure*}

The data were correlated using the VLBA-DiFX implementation of the software correlator developed by \citet{Del07}, and subsequently reduced according to standard procedures within {\sc AIPS}.  We corrected the correlator model for the updated Earth orientation parameters and ionospheric effects, using the system temperatures to calibrate the amplitude scale before deriving instrumental phase solutions using the fringe finder source.  Fringe fitting was then performed on the phase reference source (the brighter source J1748-2907 where available, otherwise J1740-2929), which was subjected to iterative imaging and self-calibration.  The final image of the phase reference calibrator was then used as a model for bandpass calibration before transferring bandpass, amplitude and phase solutions to H1743-322.  However, the large calibrator throw (Table~\ref{tab:calibrators}) coupled with the low elevation of the sources meant that unsolved phase gradients compromised the astrometric accuracy of the observations, with the positions of both H1743-322 and the check source being found to vary between epochs.  Since this had a detrimental effect on our ability to measure the proper motions of the jet ejecta, we tested a hybrid approach which was found to give a much more stable position of the core radio emission from H1743-322 during the LHS in the initial phase of the outburst.  We first solved for the short timescale phase variations using the original phase reference source.  Having transferred the solutions to H1743-322, we then re-derived the phase solutions using the check source J1744-3116 as the phase calibrator.  Owing to the infrequent observations of this source, this imposed a longer-timescale phase variation on the solutions.  The important effect was to re-reference the astrometry to provide a more stable target position relative to the assumed position of the closer calibrator J1744-3116.

Owing to the strong scattering along the line of sight, we were restricted to baselines shorter than 25\,M$\lambda$ at 8.4\,GHz, implying that we could only derive meaningful information from baselines between the five south-western antennas (FD, KP, LA, PT, OV).  The loss of half our antennas both reduced our sensitivity by a factor of 2 relative to that predicted for the full array, and degraded our angular resolution and {\it uv}-coverage.  Nevertheless, we were able to image the data, using natural weighting to maximize our sensitivity.  The results are summarized in Table~\ref{tab:vlba}.

\subsection{ATCA}

Between 2009 May 27 (MJD 54978.8) and 2009 August 7 (MJD 55050.34), we performed a total of 8 observations of H1743-322 with the Australia Telescope Compact Array (ATCA). The observations were carried out using the new Compact Array Broadband Backend \citep[CABB;][]{Wil11}. This upgrade has provided a new broadband backend system for the ATCA, increasing the maximum bandwidth from 128 to 2048\,MHz. Each observation was conducted simultaneously in two different frequency bands, with typical central frequencies of 5.5 GHz and 9 GHz. Various array configurations were used during our observing campaign. On May 27 and July 17 the array was in the compact, hybrid (north-south and east-west baselines) configurations H214 and H75 respectively. The more extended configurations 6A and 6D were used for the remaining 6 observations.

The ATCA has orthogonal linearly polarized feeds and full Stokes parameters (I, Q, U, V) were recorded at each frequency.  We used PKS 1934--638 for absolute flux and bandpass calibration, and PMN 1729--37 to calibrate the antenna gains and phases as a function of time.  We determined the polarization leakages using either the primary or the secondary calibrator, depending on the parallactic angle coverage of the secondary. Imaging was carried out using a multi-frequency clean algorithm \citep{Sau94} to take into account spectral variations of the source flux density across the 2 GHz bandwidth. The editing, calibration, Fourier transformation, deconvolution and image analysis were carried out with the Multichannel Image Reconstruction, Image Analysis and Display ({\sc miriad}) software \citep*{Sau95}.  The results of the ATCA observations are given in Table~\ref{tab:atca}.

\begin{table*}
\begin{center}
\small
\begin{tabular}{lclccc}
\hline\hline
Date & MJD$^a$ & X-ray & Frequency & Flux density & $\alpha$\\
& (d) & state$^b$ & (GHz) & (mJy\perbeam)$^c$ &\\
\hline
2009 May 27 & 54978.8 & $-$ & 5.50 & $2.40\pm0.07$ & $-0.28\pm0.08$\\
 & & $-$ & 8.62 & $2.12\pm0.05$ & $-$\\
2009 Jun 4 & 54986.6 & HIMS to SIMS & 5.50 & $0.26\pm0.02$ & $-0.73\pm0.31$\\
 & & $-$ & 8.80 & $0.18\pm0.03$ & $-$\\
2009 Jun 5 & 54987.6 & SIMS to HSS & 5.50 & $23.82\pm0.14$ & $0.03\pm0.02$\\
 & & $-$ & 9.00 & $24.20\pm0.11$ & $-$\\
2009 Jun 6 & 54988.8 & HSS & 5.50 & $24.37\pm0.09$ & $0.15\pm0.01$\\
 & & $-$ & 9.00 & $26.19\pm0.16$ & $-$\\
2009 Jun 17 & 54999.8 & HSS & 5.50 & $0.23\pm0.03$ & $-0.66\pm0.45$\\
 & & $-$ & 9.00 & $0.17\pm0.03$ & $-$\\
2009 Jun 20 & 55002.6 & HSS & 5.50 & $0.20\pm0.02$ & $-0.67\pm0.48$\\
 & & $-$ & 9.00 & $0.14\pm0.03$ & $-$\\
2009 Jul 13 & 55025.4 & HIMS to LHS & 5.50 & $<0.16$ & $-$\\
 & & $-$ & 9.00 & $<0.30$ & $-$\\
2009 Aug 7 & 55050.8 & LHS & 5.50 & $0.10\pm0.02$ & $-$\\
 & & & 9.00 & $<0.11$ & $-$\\
\hline
\end{tabular}
\end{center}
{\caption{\label{tab:atca}Log of the ATCA observations, showing for each epoch the X-ray state, the measured flux densities, and the derived radio spectral index of H1743-322.
\newline \newline
$^a$ The reported MJD corresponds to the midpoint of the observations.
\newline
$^b$ X-ray states are those derived by \citet{Mot10}.  Where the radio data were taken between two X-ray observations in different states, we report the relevant transition.
\newline
$^c$ In the case of non-detections, we quote the $3\sigma$ upper limit to the source brightness.
}}
\end{table*}

\subsection{RXTE}
\label{sec:x-rays}

We analysed all publicly available observations of the 2003 and 2009 outbursts of H1743-322 in the {\it RXTE} archive.  The data were reduced using the {\sc heasoft} software package v6.8, following the standard steps described in the {\it RXTE} cookbook\footnote{http://heasarc.gsfc.nasa.gov/docs/xte/data\_analysis.html}.  We constructed Hardness Intensity Diagrams (HIDs) from  the Proportional Counter Array \citep[PCA;][]{Jah06} data.  We produced background subtracted light curves binned at 16\,s using PCA Standard 2 mode data from Proportional Counter Unit (PCU) 2 (all layers). The light curves were divided into three energy bands, 2.1--4.9, 8.6--18.3 and 2.1--18.3 keV (according to the correspondence between fixed energy channels and energy valid for PCA gain epoch 5). We defined the hardness ratio (HR) as the ratio of count rate in the bands 8.6--18.3 keV and 2.1--4.9 keV and the intensity as the count rate in the energy band 2.1--18.3 keV. 

All the observations we used fall in PCA gain epoch 5 (from 2000 May 13 until the present). However, (small) continuous gain changes are reported within one epoch. To correct for this effect, which could affect our comparison of the 2003 and 2009 outbursts, we estimated the variation of the HR of the Crab between 2003 and 2009. Because the spectrum of the Crab is supposed to be constant, any changes observed in the hardness of the Crab are most probably caused by changes in the instrumental response. We selected RXTE observations covering the same period as our observations of H1743-322. We estimated an increase of $\sim 5.7$ per cent in the average HR of the Crab from 2003 to 2009. Consequently, we divided all the HRs calculated for H1743-322 in 2009 by a factor 1.057 to match the 2003 data. The changes in the PCA gain, although small, might also affect the source count rate. On average, the Crab count rate in the 2.1--18.3 keV decreased by $\sim 4.6$ per cent from 2003 to 2009. We thus corrected the intensities of H1743-322 to account for this change. 

To determine the spectral state of the source at the time of each RXTE observation, we used the classifications of \citet{Mot10}.  These were based on timing properties (the fractional rms variability, the presence of QPOs, and the overall shape of the power density spectra), hardness ratios, and also on the relative contributions of disc and power-law components to the X-ray spectra (see for example their fig.~5).  This latter diagnostic was important in distinguishing the SIMS (with significant power-law contribution) from the HSS (dominated by the thermal disc emission), since the fractional rms variability and hardness ratios could be very similar in these two states (see Fig.~\ref{fig:h1743_lcs}).

\section{Results}

\subsection{Radio light curves and spectra}

\begin{figure}
\centering
\includegraphics[width=\columnwidth]{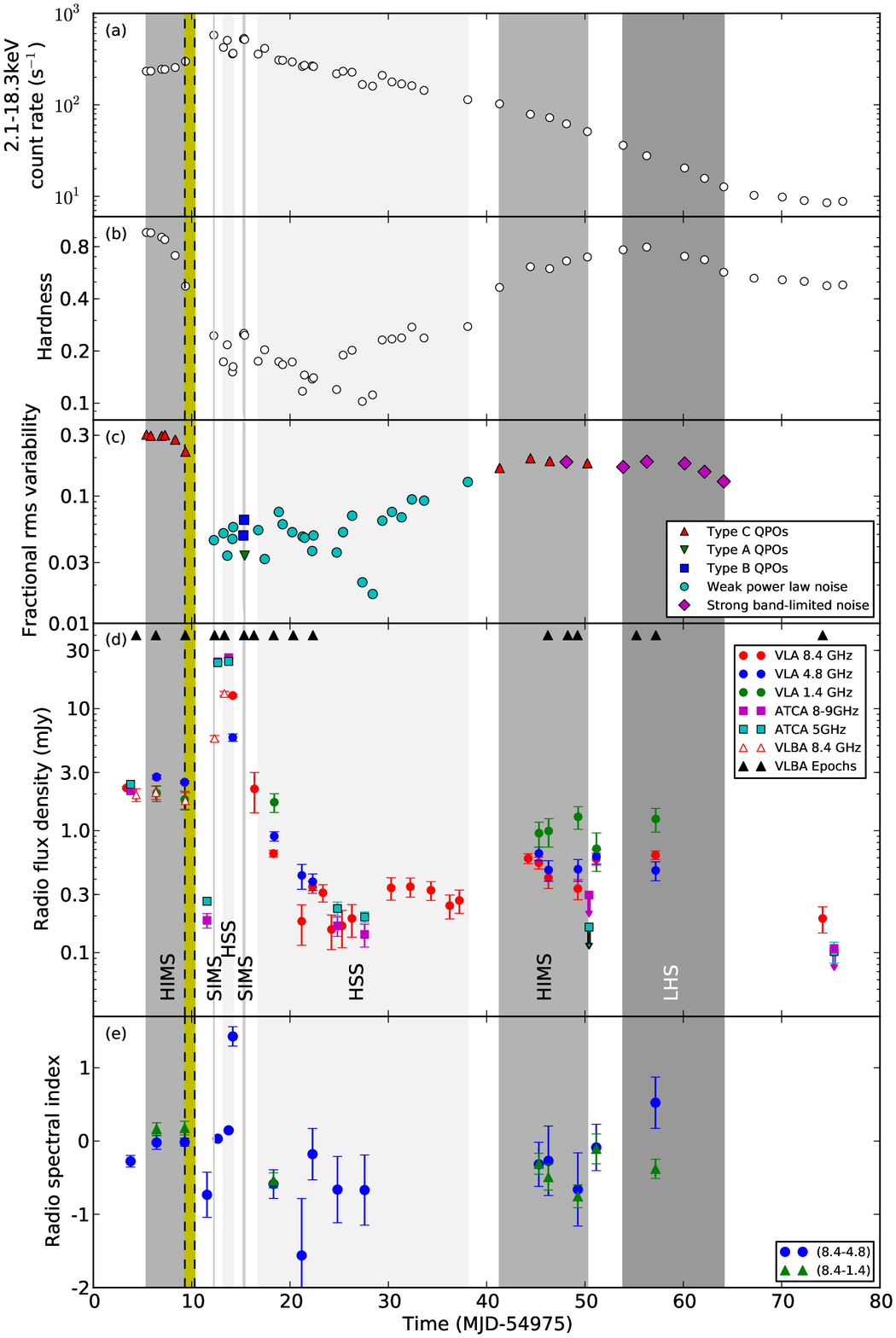}
\caption{Evolution of the 2009 outburst of H1743-322.  (a) 2.1--18.3\,keV count rate as measured by {\it RXTE}/PCA.  (b) X-ray hardness, defined as the ratio of the count rate in the 8.6--18.3\,keV band to that in the 2.1--4.9\,keV band.  (c) X-ray timing information.  The plot shows the integrated fractional rms variability in the 0.1--64\,Hz band, with symbols representing notable features in the power spectrum.  Data taken from \citet{Mot10}.  (d) radio light curves.  (e) radio spectral indices.  The grey shaded bands show the X-ray states determined for each {\it RXTE} observation by \citet{Mot10}, as labelled in panel (d).  The yellow band (bordered by the vertical dashed lines) shows our inferred ejection date for the relativistically-moving knots (see Section~\ref{sec:ejection}).}
\label{fig:h1743_lcs}
\end{figure}

The radio light curves and the evolution with time of the radio spectral index $\alpha$ (defined as $S_{\nu}\propto\nu^{\alpha}$) are shown in Fig.~\ref{fig:h1743_lcs}.  Also plotted for comparison purposes are the X-ray count rate, hardness and integrated fractional rms variability from {\it RXTE}/PCA monitoring, together with the X-ray state classifications of \citet{Mot10}.  As expected \citep[e.g][]{Fen04,Fen09}, changes in the radio emission are more closely coupled with X-ray spectral and timing information than with the overall X-ray light curve.

By the time we triggered our observing campaign, the source was already in the HIMS, with a steady radio flux density of $\sim2$\,mJy and a flat radio spectrum.  This persisted until the end of the HIMS, after which the ATCA detected a quenching of the radio emission and a steepening of the radio spectrum to $\alpha=-0.76\pm0.32$ on MJD\,54986.6.  Quenching of the radio emission prior to a major ejection event is a well-known phenomenon in Cygnus X-3 \citep{Wal94}, and has also been observed in XTE J1859+226 \citep{Bro02}, associated with a brief steepening of the radio spectrum.  It is likely that this observation corresponds to the phase of instability in the jet prior to the main outburst noted by \citet{Fen04}.

Within 0.8\,d of the ATCA detection of quenched radio emission, the VLBA detected the onset of the radio flare, on MJD\,54987.3.  This observation was almost simultaneous with the first X-ray detection of the SIMS.  The peak of the radio outburst, as detected with the ATCA, reached 26.2\,mJy on MJD\,54988.8, and the radio brightness as measured by all instruments then decreased for the next 14\,d as the source moved into the HSS.  At the peak of the flare, the radio spectrum was slightly inverted, consistent with expanding synchrotron plasma that was self-absorbed at lower frequencies, with a turnover in the observed ATCA frequency range.

While {\it RXTE} detected an excursion from the HSS back into the SIMS on MJD\,54990.3, the radio sampling was too coarse to detect a temporary rebrightening at this time.  The VLBA did not detect the source on MJD\,54990.3, and the measured VLA radio brightness on MJD\,54991.4 was consistent with the decay from the original flare.  The only evidence for a second ejection event is the inverted spectrum between 4.9 and 8.5\,GHz on MJD\,54989.2 ($\alpha=1.4\pm0.2$).  Given that these data were taken after the peak of the main flare as measured by ATCA, this suggests the early stages of a second ejection event.

Following this VLA detection of an inverted spectrum, the radio spectra were steep during the HSS.  The source was detected throughout this phase of the outburst, with a slight increase in the brightness at 8.5\,GHz on MJD\,55005.3, following which the radio emission was relatively stable at a level of $\sim0.3$\,mJy.  The unified model of \citet{Fen04} suggests that the compact core jet should be suppressed in this phase, so one likely explanation for the low-level emission observed at late times in the HSS is an ongoing interaction between the jet and the interstellar medium through which it was propagating \citep[as suggested by][to explain the residual radio emission detected in the HSS of XTE J1650-500]{Cor04}.  However, the source was not detected with the VLBA during this period, and the relatively compact configurations of the VLA and the ATCA did not provide sufficient resolution to distinguish between any such external shocks and a compact core jet.  Thus in the absence of either spectral or imaging information, we cannot verify this hypothesis.

After the reverse transition to the HIMS, the radio brightness showed a second slight increase, to $\sim0.6$\,mJy, with a spectrum consistent with being flat.  We interpret this as the reignition of the compact core jet.  As the X-ray emission decayed late in the outburst, the radio brightness also decreased, reaching 0.1\,mJy by MJD\,55050.3.

\subsection{Radio imaging}
\label{sec:imaging}

H1743-322 was detected by the VLBA on 5 occasions during the outburst (Table~\ref{tab:vlba}).  Three of these were during the HIMS prior to the radio flare, and the following two were during the rise phase and at the peak of the flare.  The resulting images, together with the stacked image from all data during the outburst decay, are shown in Fig.~\ref{fig:h1743_8ghz}.  In the first three observations during the HIMS, we detected only the compact core jet, which was unresolved down to a beam size of $14.8\times5.6$\,mas in P.A. 6.6\degree\ E of N.  In contrast, at the peak of the flare, we observed two distinct components, suggesting a two-sided ejection event.  We note that the weak component seen at the eastern edge of the VLBA image from 2009 June 5 (with relative co-ordinates (65,-23)\,mas in R.A.\ and Dec., respectively) has a significance of $<5\sigma$. Given the marginal nature of this detection, and since its orientation is not aligned with the observed axis of the bright jet components, we do not believe it to be real.

\subsubsection{No evidence for diffuse, extended emission}

In the initial LHS, the flux density measured with the VLBA was at a level consistent (at the 2$\sigma$-level) with that observed during the quasi-simultaneous VLA and ATCA observations.  However, the two VLBA observations during the rise and at the peak of the flare recovered only 25--50 per cent of the integrated flux density measured a few hours later by the ATCA.  Although this could indicate excess diffuse emission that was resolved out by the VLBA, it is likely that the majority of the discrepancy can be attributed to either timing effects during the rise phase of the flare, or to atmospheric phase decorrelation.  The VLBA observation on MJD $54987.33\pm0.10$ was taken during the rise phase of the flare, 6 hours prior to the closest ATCA observation.  The ATCA data (Table \ref{tab:atca}) show that the 9-GHz radio emission increased from 0.2 to 24.2\,mJy over the 24-hour period from MJD 54986.6 to 54987.6, so a 6-hour time offset could correspond to a significant change in flux density.

To test the effects of phase decorrelation, we compared the flux density recovered from the calibrator sources J1748-2907 and J1740-2929 after applying the phase corrections derived using J1744-3116, with that derived from self-calibration of those same two sources.  This showed that phase decorrelation could reduce the recovered flux density by up to 50 per cent on MJD 54987.3, and by up to 25 per cent on MJD 54988.3 (when the atmosphere was presumably somewhat more stable).  Taking into account the effects of both time offsets and atmospheric phase decorrelation, we conclude that although we cannot rule out the existence of diffuse emission that is resolved out by the VLBA, such a component is not required to reconcile the measured VLBA fluxes with those seen at the ATCA.

\subsubsection{No evidence for free-free absorption in the inner regions}

Given the high inclination of the system and the presence of a disk wind \citep{Mil06}, we should address whether there is any evidence for free-free absorption in the inner radio-emitting regions, as has been found in the case of SS\,433 \citep{Par99}.  Since the outflow rate estimated for the wind in H1743-322 \citep{Mil06} is several orders of magnitude lower than that in the equatorial wind of SS\,433 \citep{Blu01}, there should be significantly less free-free absorbing material along our line of sight to the inner jets.  This is supported by the radio spectrum, which is only ever highly inverted (as would be expected for free-free absorption) during the rise phase of the flare (Fig.~\ref{fig:h1743_lcs}), when an inverted spectrum can be attributed to low-frequency synchrotron self-absorption in the expanding ejecta.  Finally, we always detect the core of the system with our VLBA imaging during the HIMS, at a position intermediate between the eastern and western jet ejecta seen later in the outburst (Fig.~\ref{fig:h1743_8ghz}).  Thus there is no reason to believe that there is significant free-free absorption in the inner regions of the system.

\subsection{Astrometry}

\subsubsection{Core position}
\label{sec:core}

\begin{figure}
\centering
\includegraphics[width=\columnwidth]{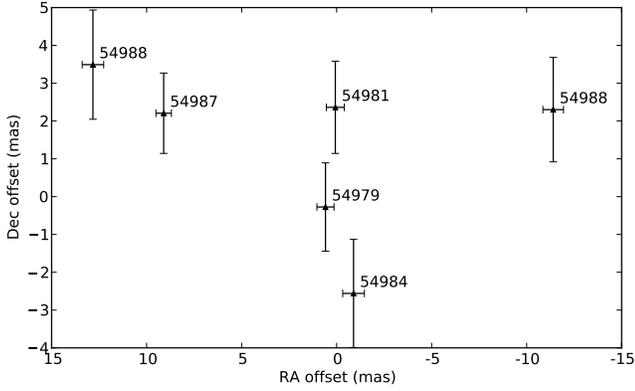}
\caption{Fitted positions of all detected VLBA components from H1743-322, relative to the weighted mean core position of 17$^{\rm h}$46$^{\rm m}$15\fs59637, $-32$\degr14\arcmin00\farcs860 (J\,2000).  East is to the left and north is up.  Each point is labelled with the MJD of the observation.  Note that the positions of both components found for the final observation (MJD 54988) are shown.}
\label{fig:position_offsets}
\end{figure}

The early triggering of the observations and consequent detection of the compact core jet provided a point of reference for the subsequent observations during the radio outburst.  We measured the source position, relative to the assumed position of the calibrator source J1744-3116 as given in Table~\ref{tab:calibrators}.  The three observations from the HIMS showed a very consistent position in R.A.\ but a scatter of $\sim5$\,mas in Dec.\ (Fig.~\ref{fig:position_offsets}).  The calibrator throws to both the short-timescale phase reference calibrators J1748-2907 and J1740-2929 and to the final calibrator source J1744-3116 are all significantly greater in Dec.\ than in R.A..  Therefore the phase gradients with elevation expected for such a low-declination source will have a greater effect on the Dec.\ component of position than on the R.A.\ component, so such a discrepancy in the scatter is not surprising.  A weighted mean of the three measured core positions from the HIMS provides our best estimate of the true position of H1743-322, 
\begin{equation}
\begin{split}
\notag
{\rm RA} &= 17^{\rm h}46^{\rm m}15\fs59637(6)\\
{\rm Dec.} &= -32\degr14^{\prime}00\farcs860(2)\qquad{\rm (J\,2000)},
\end{split}
\end{equation}
where the assumed errors are purely statistical.  To this we must add a systematic error of 2.7\,mas in R.A.\ and 8.0\,mas in Dec.\ from combining the quoted uncertainty on the calibrator position\footnote{http://www.vlba.nrao.edu/astro/calib/} and the uncertainty from the calibrator throw to J1744-3116 (using a linear extrapolation of the results of \citealt*{Pra06} to the low declination of the target source).  We note that our quoted position is measured relative to the final calibrator source, J1744-3116, whose assumed position is given in Table~\ref{tab:calibrators}.

\subsubsection{Ejection event}
\label{sec:ejection}

\begin{figure}
\centering
\includegraphics[width=\columnwidth]{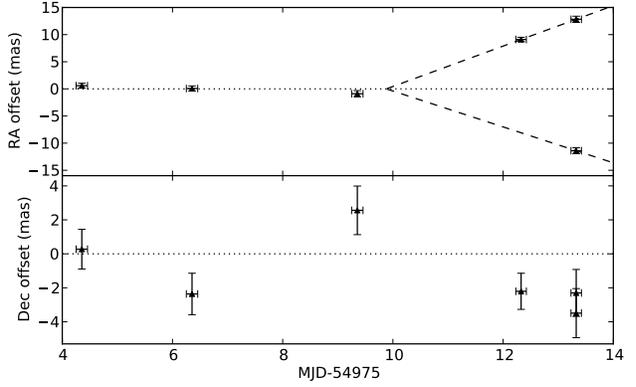}
\caption{Angular separation from core position in R.A.\ and Dec.\ as a function of time.  We also show a fit to the approaching component for epochs MJD\,54987 and 54988, using this to derive an ejection date and hence fit the proper motion of the receding component.}
\label{fig:proper_motions}
\end{figure}

While there is some scatter in the Dec.\ component of the source position during the HIMS, the R.A.\ is constant to within 1.5\,mas.  After the transition to the SIMS and the subsequent HSS, the measured source positions at both 8.4 and 15\,GHz on MJD\,54987 and 54988 show evidence of components moving away from the X-ray binary (Fig.~\ref{fig:proper_motions}).  On MJD\,54988, we see two radio-emitting components, aligned along an axis oriented $87.2\pm3.6^{\circ}$ E of N.  This is fully consistent with the jet axis ($89.0\pm1.5^{\circ}$ for the eastern jet, $-91.7\pm1.8^{\circ}$ for the western jet) determined by \citet{Cor05}, suggesting that the jet axis has not precessed since 2003.  Further, their measurement of larger angular separations for the eastern component is in agreement with our VLBA observations (the eastern component appearing first, on 2009 June 5), despite the fact that their jet knots had decelerated by the time the radio emission was observed.  This suggests that the interstellar medium surrounding the source is not sufficiently inhomogeneous that the approaching (eastern) ejecta were decelerated more than the receding (western) component in the observations of \citet{Cor05}.

The angular separation of the two components from the weighted mean core position is shown in Fig.~\ref{fig:proper_motions}.  Assuming that we are identifying a single ballistically-moving component on each side of the core (likely, given that only one day separates the two observations), a linear fit to the displacement along the jet axis of the approaching (eastern) component on MJD 54987 and 54988 gives a proper motion $\mu_{\rm app}=3.7\pm0.7$\,mas\,d$^{-1}$ (an apparent velocity of $v_{\rm app}=0.18c$ at 8.5\,kpc).  This fit determines the ejection date as $t_{\rm ej} = $MJD\,54984.9$^{+0.4}_{-0.6}$.  While this ejection date was derived from only two measurements of the ejecta positions, the agreement found between the positions measured in the 15\,GHz and 8.4\,GHz VLBA images gives us confidence in our estimate.  With only two data points, there may be additional systematic uncertainties affecting the determination of the ejection date, although their magnitude is not possible to quantify from our data.  While we cannot definitively rule out deceleration of the ejecta on such small angular scales \citep[e.g.][]{Yan10}, we consider it unlikely, since the ejecta from the 2003 outburst decelerated much further downstream.  The continued jet activity of H1743-322 since that time (Section \ref{sec:h1743}) makes it unlikely that the density of the interstellar medium close to the central binary could have been substantially increased in the interim.

Assuming therefore that our derived ejection date is correct, then a ballistically-moving receding (western) component would have a proper motion $\mu_{\rm rec}=3.3\pm0.8$\,mas\,d$^{-1}$ ($v_{\rm app}=0.16c$ at 8.5\,kpc).  The proper motions of both approaching and receding jet knots are therefore significantly lower than those derived by \citet{Cor05} on larger angular scales ($21.2\pm1.4$\,mas\,d$^{-1}$ for the approaching jet, $13.3\pm0.6$\,mas\,d$^{-1}$ for the receding jet, although the jets had likely decelerated by this point; see also \citealt{Hao09}).  Either the jet speed varies between outbursts (possibly correlated with jet power; the 2003 outburst was significantly brighter in the radio band than the one presented here; see \citealt{McC09}), the ballistic motion model is not applicable to our data, or we are seeing different components in the eastern jet in the two images.

\section{X-ray diagnostics of the ejection event}

A major goal in the study of the coupling between inflow and outflow in X-ray binaries is to relate the observed changes in the radio emission (probing the relativistic jets) to those seen in the X-ray band (probing the inner regions of the accretion flow, and possibly the base of the jets).  The so-called `jet line' proposed by \citet{Fen04} to mark the transition from steady, partially self-absorbed, compact core jets to bright, transient, relativistically moving ejecta, is a useful way to visualise this transition.  However, as discussed by \citet{Fen09}, the reality is likely to be more complex.  Nevertheless, as we now discuss, even if the exact sequence of events has not been uniquely identified to date, certain characteristic X-ray spectral and timing indicators appear to be closely coupled with the radio ejection event.

\subsection{Spectral indicators}

\begin{figure}
\centering
\includegraphics[width=\columnwidth]{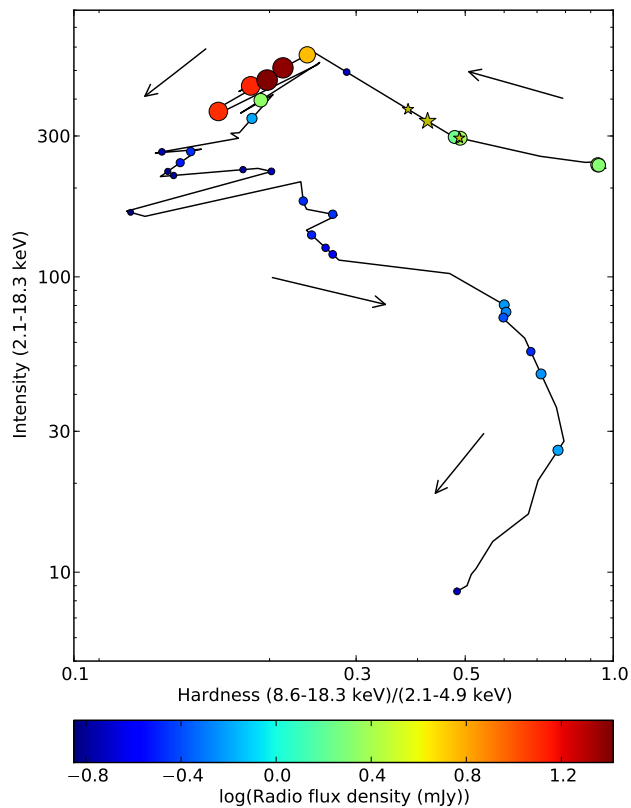}
\caption{Hardness-intensity diagram for the 2009 outburst of H1743-322.  Circles indicate radio detections, with the size and shade of the circles indicating the measured radio flux density.  The location in the HID at the time of each radio observation has been interpolated from {\it RXTE} PCA observations within 2\,d.  The large star represents the moment of ejection as estimated in Section~\ref{sec:ejection}, with the smaller stars showing the extent of the $1\sigma$ error bars on this date.  Arrows show how the source evolves with time.  The source follows a fairly canonical path through the HID.}
\label{fig:hid}
\end{figure}

The evolution of black hole X-ray binary outbursts is often visualized in an X-ray hardness-intensity diagram (HID).  This shows how X-ray intensity variations couple with X-ray spectral shape, with harder spectra being dominated by the power-law component and softer spectra by blackbody disc emission.  As shown in Fig.~\ref{fig:hid}, H1743-322 underwent a fairly typical evolution through the HID (compare with the schematic model in fig.~7 of \citealt{Fen04} and the observed behaviour of the 14 outbursts plotted in fig.~1 of \citealt{Fen09}).  We have overplotted the radio detections, showing the variation of the radio brightness as the source evolved through the outburst.  There is clear evidence for radio flaring as the source moves left across the top of the diagram from the HIMS to the HSS.  Also noteworthy are the position of the radio quench phase and the radio rebrightening as the source moves back to a harder state towards the end of the outburst.

\subsection{Variability indicators}

The transition from the HIMS to the SIMS is defined by X-ray timing information, namely a sharp decrease in the integrated fractional rms variability and the possible appearance of Type B QPOs \citep{Bel05}.  Using case studies of the three sources XTE J1550-564, XTE J1859+226 and GX339-4, \citet{Fen09} considered whether the sharp changes in X-ray power spectra seen during the transition from the HIMS to the SIMS could be used to determine the exact moment of the major ejection event, as previously proposed by \citet{Fer06} and \citet{Kle08}.  However, \citet{Fen09} deemed the available radio coverage insufficient to associate the exact moment of ejection with any particular X-ray spectral or timing characteristics.  Sparse time sampling, coupled with the (unknown) intrinsic delay between ejection and radio flaring caused by both optical depth effects and the time taken for the development of the hypothesised internal shocks, made it impossible to determine the moment of ejection and hence search for any corresponding X-ray signature.  With our dense radio coverage, the core location provided by the VLBI detection in the early phase of the outburst (Section~\ref{sec:core}), and the fitted proper motion of the ejecta (Section~\ref{sec:ejection}), we finally have a data set in which it is possible to probe this connection directly.

\begin{figure}
\centering
\includegraphics[width=\columnwidth]{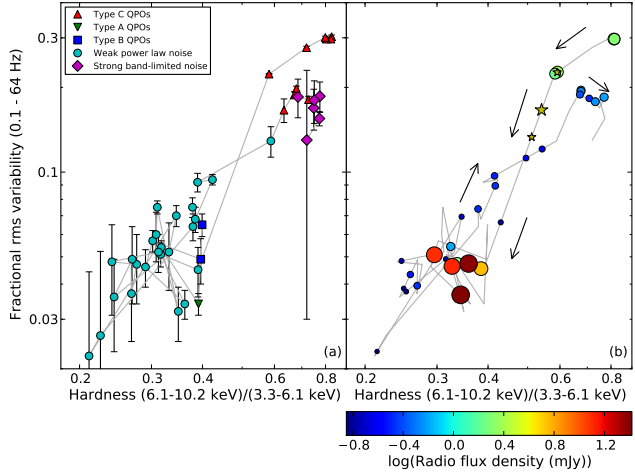}
\caption{Hardness-rms variability diagram (HRD) for the 2009 outburst of H1743-322.  (a) Measured X-ray data with error bars, with symbols representing the features in the power spectrum.  Data taken from \citet{Mot10}. While we note their use of different bands for calculating the hardness ratio from those defined in Section~\ref{sec:x-rays}, this should not affect the overall picture. (b) Variation of radio flux density with position in HRD.  X-ray spectral and timing information at the time of each radio observation have been interpolated from {\it RXTE} PCA observations within 2\,d. The size and shade of the circles indicate the measured radio flux density.  The large star represents the moment of ejection as estimated in Section~\ref{sec:ejection}, with the smaller stars showing the extent of the $1\sigma$ error bars on this date.  Arrows show how the source moves along the grey trace through the diagram with time.  The moment of ejection appears to coincide well with the drop in rms variability.}
\label{fig:hrd}
\end{figure}

Fig.~\ref{fig:hrd} shows how the X-ray variability properties vary with spectral hardness during the 2009 outburst of H1743-322, and also how the radio emission relates to these diagnostic X-ray characteristics.  Our inferred ejection event occurs as the fractional rms variability begins to drop off sharply, coinciding with the disappearance of Type C QPOs.  Unfortunately, this phase of the outburst coincides with a 3\,d gap in the {\it RXTE} coverage.  Although we do not analyse {\it Swift} data in this paper, we note that there were no {\it Swift} observations during the gap in {\it RXTE} coverage, so we cannot constrain the simultaneity of the ejection event and this X-ray signature more accurately. The radio quenching occurs as the X-ray variability decreases, but the time taken for the development of shocks and their evolution to become optically thin in the GHz regime means that the peak radio emission is seen only in the zone of reduced variability.

\section{Jet properties}

\subsection{Jet speeds}

Special relativistic effects imply that the ratio of angular separations of approaching and receding components from the core at any given epoch is
\begin{equation}
\frac{l_{\rm app}}{l_{\rm rec}} = \frac{1+\beta\cos\theta}{1-\beta\cos\theta},
\end{equation}
where $\beta c$ is the jet speed and $\theta$ is the inclination angle to the line of sight \citep[e.g.][]{Mir99}.  From the epoch in which we detect both approaching and receding components, our best core position implies $l_{\rm app}/l_{\rm rec}=1.12\pm0.13$, which gives $\beta\cos\theta=0.058\pm0.056$.  Combined with the inclination angle of $75\pm3$\degree\ derived by \citet{Ste12}, this implies $0.19<\beta<0.28$.

As noted in Section~\ref{sec:ejection}, the proper motions measured during the 2009 outburst are significantly lower than those measured (albeit on larger scales) in the 2003 outburst, leading to lower derived jet speeds.  \citet{Cor05} found a jet speed $\beta=0.79$ (with $\theta=73$\degree) for the 2003 outburst, assuming a distance of 8\,kpc and ballistic motion of the approaching and receding ejecta.  Even in the case of decelerating jet knots, they were able to constrain $\beta>0.57\pm0.05$.  Since acceleration of the ejecta during propagation is unlikely, and since the consistent position angle of the jet axis on the plane of the sky argues against a precessing jet, it seems that the jet velocity was very different in these two outbursts.  Since few X-ray binaries have been observed with VLBI during multiple ejection events, we do not have good statistics on whether jet speeds are likely to vary between different events in a single source, although plausible evidence for a variable jet speed in GRS\,1915+105 has been found \citep{Mil07}.

\subsection{Suppression of the core jets in the soft state}

The compact core jets are known to be suppressed in the HSS of H1743-322, by a factor of $\sim700$ \citep{Cor11}.  However, our VLA observations during the HSS show low-level emission at 8.4\,GHz, with a flux density of up to $0.35\pm0.06$\,mJy.  \citet{Cor04} put forward two possible explanations to explain similar residual emission detected in the HSS of XTE J1650-500; namely a residual compact core jet, and interactions of the ejecta from the outburst with the surrounding interstellar medium.  A compact core jet should have a flat radio spectrum \citep{Bla79,Fen01}, whereas we would expect optically-thin synchrotron emission from a downstream external shock, giving rise to a steep spectrum \citep[e.g.][]{Cor02,Cor05}.  Unfortunately, we do not have any spectral information in this phase to distinguish between these two possibilities, since the source was sufficiently faint that we could not split the observing time between multiple frequencies.  However, we can stack our VLBA observations from this phase (MJD\,54991--54997) to place a $3\sigma$ upper limit of 0.27\,mJy\perbeam\ at our derived core position.  Given the uncertainties on the HSS flux densities measured by the VLA, this is insufficient to conclusively rule out a low-level compact jet, but, together with the known radio suppression in the HSS of this system \citep{Cor11}, appears to favour the downstream external shock scenario.

\section{The reverse transition}

\subsection{Jet reactivation}
\label{sec:reactivation}

Owing in part to the speed of the X-ray evolution, there is relatively little existing radio data constraining the reactivation of the jet during the reverse transition from the HSS through the HIMS to the LHS.  From a comparison of data from two different outbursts of GX339-4, \citet{Fen09} found radio upper limits during the reverse transition at the same hardness where the peak of the flare was observed during the original hard to soft transition, implying that either the jet line was not vertical or that it was not consistent between different outbursts (see Section~\ref{sec:comparison}).

The reactivation of the jet has been reported from infrared/optical monitoring of three black hole X-ray binaries \citep{Jai01,Bux04,Cor09,Rus10}. Here, the higher energy emission from the compact core jet originates in a region close to the black hole, within a light second \citep[e.g.][]{Cas10,Gan10}, and cannot be caused by colliding ejecta launched previously (assuming a single power law with $\alpha = -0.6$, an extrapolation of the brightest optically thin radio ejecta of H1743-322 would be $< 0.1$ mJy at near-infrared (NIR) frequencies; much fainter than the observed NIR jet emission in other black hole X-ray binaries at similar distances). In all cases, the optical/NIR jet became visible only after the source had fully completed the transition to the LHS. The radio emission, at least in one of these systems, brightened and peaked before the optical/NIR synchrotron emission \citep[see fig. 5 of][]{Cor09}, suggesting that the jet power (which is sensitive to the high energy emission) takes time (days--weeks) to build up after the transition from HSS to LHS \citep[e.g.][]{Rus10}.

\citet{Kal06} also argued that the compact jet did not turn back on until well after the transition back to the intermediate state, and suggested that the formation of the compact jet may require the source to be in the LHS.  However, \citet{Fen09} noted the radio reactivation during the 2003 outburst of XTE J1720-318 \citep{Bro05}, which occurred marginally prior to the source reaching the hardest spectral state.  Our data provide better constraints on the reverse transition, clearly showing a radio rebrightening between the last observation in the HSS and the first in the HIMS, 6--9\,d before the source reached the canonical LHS.

Since we detect radio emission in the HSS which we attribute to external shocks as the jets interact with the surrounding medium, we must address the question of whether this rebrightening is due to further external shocks, or to the reignition of the compact core jet.  From Table~\ref{tab:vla}, we see that the error bars on the radio spectral indices are too large to accurately discriminate between a flat spectrum compact jet and optically thin ejecta.  However, since there is no further jump in radio brightness between the first detection in the HIMS and the LHS, when we believe the compact jet should be present, this suggests that the radio brightening seen between the HSS and the HIMS is indeed due to the reactivation of the core jet.

The compact core jet is not detected in the individual VLBA data sets during the reverse transition.  Stacking the VLBA data from the HIMS and LHS during the reverse transition shows a $4.5\sigma$ detection ($0.36\pm0.08$\,mJy) close to the measured position of the core (see Fig.~\ref{fig:h1743_8ghz}).  However, the marginal nature of the detection and the positional offset ($22\pm3$\,mas in PA $-14.8\pm2.9$\degree\ E of N) imply that we cannot reliably associate this source with the compact core jet.  The proper motion of H1743-322 is unknown, but from a comparison of our measured position with that determined from VLBA observations made in 2004, it is unlikely to be large enough to have shifted the X-ray binary position 22\,mas in a month.  Since the jet axis seen by the VLBA at the peak of the outburst agrees so well with that observed several years earlier by \citet{Cor05}, we can rule out precession of the jets.  Combined with the marginal nature of the detection, its position angle (almost perpendicular to the jet axis, in the absence of significant systemic proper motion) makes it probable that this detection is simply a noise spike.

Given the noise level of 0.08\,mJy\perbeam\ in our stacked image, we might have expected to detect the compact core jet, given the flux densities of 0.41--0.63\,mJy measured by the VLA over this period.  Our VLBA non-detection might therefore suggest that the emission measured by the VLA was sufficiently extended to be resolved out by the VLBA, and might therefore be attributed to diffuse emission downstream in the jets.  However, the phase decorrelation of up to 50 per cent reported in Section \ref{sec:imaging} would reduce the source flux density below the detection threshold in the stacked VLBA image, so the VLBA non-detection cannot be used as evidence for extended emission downstream in the jets.

Assuming, therefore, that we are indeed seeing the reactivation of the compact core jet at the transition from the HSS to the HIMS, this re-ignition of the radio emission occurs well before the optical/NIR jet is believed to switch on at the transition to the LHS.  A similar sequence of events was also seen by \citet{Cor09} during the soft-to-hard transition in GX339-4.  We can contrast this with the known behaviour during the hard-to-soft transition.  \citet{Rus07} showed that the optical/NIR jets are suppressed at the top-right hand corner of the HID, as the X-ray emission begins to soften, while the radio flaring event and compact jet quenching occurs 7--12\,d later, at the hypothesised `jet line'.  While we have no optical/NIR coverage of the 2009 event in H1743-322, the radio quenching and subsequent jet ejection did occur part-way through the X-ray spectral softening (see Fig.~\ref{fig:hid}).

Thus, it seems that the quenching and reactivation of the jets at the beginning and end of an outburst are gradual processes, occurring at different times for the radio and optical/NIR jets.  Any discussion of these phenomena should therefore be very clear as to the wavelength at which the jets are being observed.  A possible explanation for the observed sequence of events could be that the jet power evolves over the course of several days, with the spectral break from optically thick to thin moving to lower/higher frequencies with time at the beginning/end of the outburst (the frequency of the spectral break determines the total radiative power of the jets).  An alternative explanation could be that it takes time for the suppression of the flow in the inner regions to propagate outwards to the surface of optical depth unity at radio frequencies.  However, since the compact, hard-state jets are not resolved in VLBI observations (see Fig.~\ref{fig:h1743_8ghz}), this would imply bulk flow speeds below $0.1c$, hence we do not favour this explanation.

\subsection{Distance}
\label{sec:distance}

\citet{Mac03} has shown that the reverse transition from HSS to LHS occurs at a fixed fraction (0.01--0.04) of the Eddington luminosity, $L_{\rm Edd}$.  A more detailed study by \citet{Dun10} showed that when using the fraction of the bolometric luminosity in the power-law component to identify the state transitions, the reverse transition occurred at 0.005--0.1$L_{\rm Edd}$, with a mean of $0.028L_{\rm Edd}$.  As shown in Section~\ref{sec:comparison}, the reverse transition for H1743-322 occurred at approximately the same 2.1--18.3\,keV X-ray intensity in both the 2003 and 2009 outbursts.  This level corresponds to a 2--20\,keV flux of $1.5\times10^{-9}$\,erg\,cm$^{-2}$\,s$^{-1}$ \citep[observation number 158 of][]{McC09}.  Using the tabulated photon index of 2.1 appropriate for that observation, we calculate the 0.5\,keV--10\,MeV flux \citep[an approximation to the bolometric flux; see][]{Mac03} to be $5.2\times10^{-9}$\,erg\,cm$^{-2}$\,s$^{-1}$.  Assuming that this corresponds to $0.028^{+0.079}_{-0.021}L_{\rm Edd}$ \citep{Dun10}, we derive a distance to the source of $7.6^{+7.3}_{-3.8}(M_{\rm BH}/10M_{\odot})^{1/2}$\,kpc, where $M_{\rm BH}$ is the black hole mass.  Using instead the range derived by \citet{Mac03} gives $6.5^{+2.7}_{-1.9}(M_{\rm BH}/10M_{\odot})^{1/2}$\,kpc. While relatively poorly constrained, these estimates are consistent with the distance measured by \citet{Ste12}, and thus imply a relatively high-mass compact object.

\section{Comparison with the 2003 outburst}
\label{sec:comparison}

\begin{figure*}
\centering
\includegraphics[width=\textwidth]{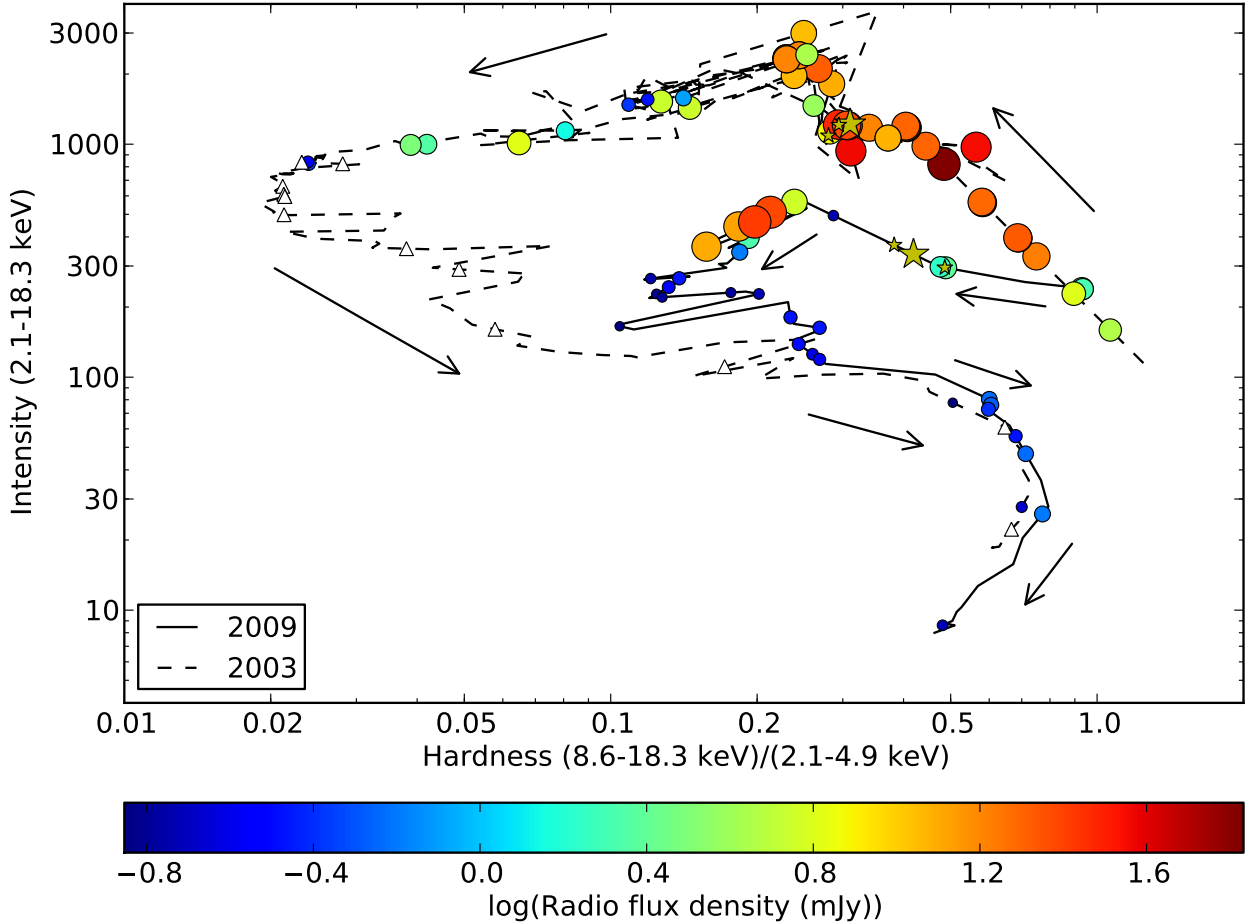}
\caption{Comparison of the hardness-intensity diagrams for the 2003 (dashed line) and 2009 (solid line) outbursts of H1743-322.  Circles indicate radio detections \citep[with measurements from the 2003 outburst taken from][]{McC09}, with the size and shade of the circles indicating the measured radio flux density.  Open triangles indicate non-detections.  The X-ray state at the time of the radio observations has been interpolated from {\it RXTE} PCA observations within 2\,d.  The large star represents the moment of ejection as estimated in Section~\ref{sec:ejection} for the 2009 outburst and by \citet{Ste12} for one ejection event during the 2003 outburst, with the smaller stars showing the extent of the $1\sigma$ error bars on these times.  The outburst in 2009 did not reach as low a hardness or as high an intensity as that in 2003.  The ejection markers clearly show that the jet line is not at a fixed hardness.  The hardness at which the ejection event occurs varies between outbursts.}
\label{fig:hid_comparison}
\end{figure*}

Fig.~\ref{fig:hid_comparison} shows the HIDs for both the 2003 and 2009 outbursts of H1743-322 plotted on the same scale, with markers corresponding to the measured radio flux densities overlaid \citep[the radio data from 2003 have been taken from][]{McC09}.  The 2003 outburst was significantly brighter in both radio and X-ray bands, and reached a much softer, more disc-dominated X-ray spectrum in the HSS.  It was a more complex outburst, with several distinct peaks in both the radio and X-ray light curves suggesting multiple ejection events.  From their model fitting of VLA data, \citet{Ste12} determined the ejection date of the jets responsible for the downstream flaring later detected by \citet{Cor05}, which they found to be MJD $52767.6\pm1.1$.  Both this ejection event and the peak radio luminosity (occurring 30 days earlier, and presumably caused by a previous ejection) occurred at different spectral hardness values compared to the 2009 outburst, suggesting that the so-called `jet line' is not constant between outbursts.  The brightest radio emission corresponds to a harder spectrum in 2003 than in 2009.  We see no evidence for quenching of the radio jets prior to the flare in the 2003 outburst, although the cadence of the observations was not sufficiently rapid to rule out having missed such an event.

While the hard to soft transition occured at a much higher 2--18\,keV X-ray intensity in 2003, the reverse transition occured at a very similar 2--18\,keV intensity in the two outbursts.  We also see that the radio emission assumed to arise from the compact core jet reactivated at a similar point in both outbursts.  This reactivation occurred slightly prior to the hardest spectral state being reached, in what is defined by \citet{Mot10} as the HIMS, or by \citet{McC09} as a hybrid state with characteristics of both the hard and steep power-law states.  As discussed by \citet{Bel10}, such a hybrid state should correspond roughly to the HIMS.

\begin{figure}
\centering
\includegraphics[width=\columnwidth]{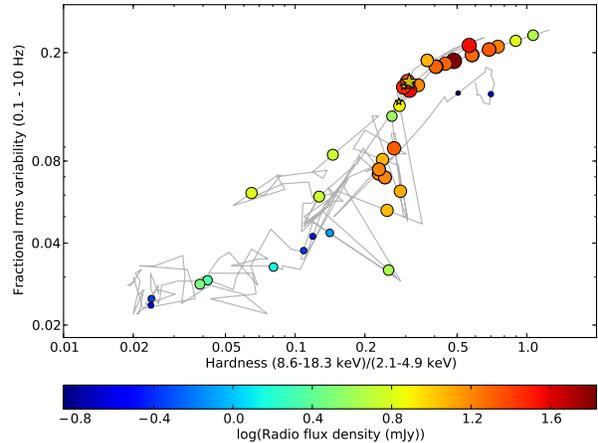}
\caption{HRD for the 2003 outburst of H1743-322.  The size and color of the markers show the radio flux density.  X-ray spectral and timing information at the time of each radio observation has been interpolated from {\it RXTE} PCA observations within 2\,d.  Timing information has been taken from \citet{McC09}.  The large star represents the moment of ejection as estimated by \citet{Ste12}, with the smaller stars showing the extent of the $1\sigma$ error bars on this date.  Despite the location of the ejection event in the HRD, it occurred as the variability was increasing; the next significant drop in the rms did not occur until 17 days later.}
\label{fig:timing_2003}
\end{figure}

The HRD for the 2003 outburst is shown in Fig.~\ref{fig:timing_2003}.  In contrast to the 2009 outburst (Fig.~\ref{fig:hrd}), the peak radio flux density occured before the sharp drop in the integrated fractional rms variability.  While we do not know the exact ejection date corresponding to the radio peak in 2003, owing to a lack of high angular resolution radio data, we know from the measurements of \citet{Cor05} that the jet speed was higher (implying a smaller delay between ejection and the formation of internal shocks should the core jet have a similar Lorentz factor in the LHS in both cases).  Assuming that the radio peak occurred after the initial ejection event, the date of the first ejection in the 2003 outburst must have occurred before the initial drop in fractional rms.

As noted above, the 2003 outburst was a complicated event, with multiple peaks in the radio light curve. \citet{Ste12} determined the ejection date associated with the third major radio peak of the outburst.  When compared with the results presented by \citet{McC09}, their derived ejection date follows a spike in the X-ray count rate, and is simultaneous with an increase in the fractional rms variability from $<5$ per cent to $\sim 15$ per cent, in stark contrast to the 2009 outburst, when we infer the moment of ejection to have been immediately before (or during) a drop in fractional rms variability (bearing in mind the 3-day gap in the X-ray coverage).  This increase in fractional rms variability also coincided with an increase in the QPO amplitude and quality factor \citep{McC09}, i.e.\ with the clear appearance of classical Type C QPOs.  If the ejection date derived by \citet{Ste12} is correct, then there is no unique X-ray variability signature that can be associated with the ejection events in both 2003 and 2009.

\section{Discussion}

The 2009 outburst of H\,1743-322 was a relatively simple event, and the lack of multiple overlapping flares provided a clean data set with which to test the phenomenological model for the disc-jet coupling in black hole X-ray binaries \citep{Fen09}.  The high angular resolution VLBA data allowed us to track back the radio ejecta to find the epoch at which the transient jets were launched, such that we do not need to rely on the epoch of peak radio flux density as a proxy for the ejection date.  We now use this information to reconsider the unified model of \citet{Fen04}, attempting to answer some of the questions raised by \citet{Fen09}.  However, we note the caveat that this is a single outburst in a single source, and that more multi-wavelength campaigns incorporating high-cadence, high angular resolution VLBI observations are required to verify our conclusions.

Within the uncertainties created by the RXTE sampling, we find that the ejection date is consistent with the transition from HIMS to SIMS, at the point where the Type C QPOs disappear and the integrated fractional rms variability begins to decrease.  This is shortly before the detection of quenched, optically-thin radio emission.

The most detailed comparison to date of the connection between radio flaring and X-ray timing signatures was presented by \citet{Fen09} (see their figs.~4 and 5).  They investigated possible connections between the radio flaring events in XTE J1550-564, XTE J1859+226 and GX339-4 and the contemporaneous drops in rms variability and appearance of Type B QPOs.  Since the exact sequence of events appeared to differ between the three sources, they concluded that there appeared to be no causal connection between timing signatures from the inner disc and jet ejection events.  However, they noted that any intrinsic signature could be blurred by both the phase of jet instability prior to a radio flare and the time delays incurred in producing optically thin radio emission from internal shocks.  In light of the conclusions drawn from our study of the 2009 outburst of H1743-322, we re-examine the flares in their three sources, considering also the 2010 flare of XTE J1752-223 and the 2003 flare of H1743-322.

\subsection{The 1998 outburst of XTE J1550-564}

The X-ray emission from the 1998 flare of XTE J1550-564 peaked on MJD 51076.0 \citep{Sob00}, followed by a radio flare peaking at 843\,MHz on MJD 51078 \citep{Han01}.  If the radio flare followed the canonical pattern of an adiabatically expanding synchrotron event, we can infer that the peak at GHz frequencies might have occurred slightly prior to this.  VLBI observations on MJD 51080.7 and 51081.7 detected several radio-emitting components.  Assuming that the two outermost components in the two images can be associated with one another, the rate of increase in angular separation was $\sim115$\,mas\,d$^{-1}$, implying a zero-separation date of MJD 51078.6, coincident with the radio flare but well after the X-ray peak.  The appearance of a third component in the second image suggests that there could have been multiple ejection events during this outburst.  Considering the additional uncertainties in interpreting the images (not being phase-referenced, the true alignment of the two VLBI images is unknown), we cannot unequivocally relate the timing of the X-ray flare and the first radio ejection event.

\citet{Fen09} show that the radio flare began as the rms variability began to drop sharply.  However, we also note a slight change in the characteristics of the Type C QPOs at the time of the X-ray flare (between MJD 51076.8 and 51077.9), moving from Type C to Type C$^{\prime}$ \citep[as defined by][with weak harmonics, a double-peaked power density and rapidly shifting QPO frequency]{Rem02}.  The QPOs reverted to Type C on MJD 51078.1, and the first Type B QPO was not observed until four weeks later, on MJD 51106.95.

\subsection{The 1999 outburst of XTE J1859+226}

The 1999 flaring event of XTE J1859+226 comprised multiple radio flares, with the first (and largest) preceded by a quench on MJD 51467.5 before peaking on MJD 51468 \citep{Bro02}.  The X-ray study of \citet{Cas04} shows a transition from Type C to Type A QPOs between MJD 51466.9 and 51467.6, coinciding with a steep drop in the rms variability (from 16.0 to 2.8 per cent) and also with the radio quenching event seen by \citet{Bro02}.  Prior to this transition, Type C QPOs had been a persistent feature of the power spectrum for at least four days.  This sequence of events appears to tie in well with what we see in H1743-322; a radio quenching coincident with the disappearance of Type C QPOs and a steep drop in the rms variability.  If the two events are indeed similar, we can infer that the radio ejection event in XTE J1859+226 likely occurred at this time, with the short delay before the onset of the radio flare arising from the time taken for internal shocks to form and become optically thin.

\subsection{The 2002 outburst of GX339-4}

As noted by \citet{Fen09}, the radio flare reported by \citet{Gal04} occurred on MJD 52408, prior to the steep drop in rms variability between MJD 52410.5 and 52411.6.  Again however, we remark on the change in character of the Type C QPOs around the time of the radio flare.  According to \citet{Bel05}, the Type C QPOs changed to Type C$^{\ast}$ between MJD 52406.7 and 52410.5, before the first observation of a Type B QPO on MJD 52411.6.  Type C$^{\ast}$ QPOs \citep{Cas04} have a broader peak, a higher frequency, and a lower rms amplitude than canonical Type C QPOs.  Thus while the Type B QPOs and the drop in rms variability appear to occur after the radio flare, it is plausible that the evolution away from the canonical Type C QPOs marks the disruption of the steady jet and the beginning of the ejection event.

\subsection{The 2010 outburst of XTE J1752-223}

Although VLBI monitoring of XTE J1752-223 was carried out during its 2010 outburst, the observed one-sided ejection event \citep{Mil11b} and the deceleration of the ejecta \citep{Yan10} prevent us from accurately determining a zero-separation time.  However, the radio light curves and spectra show evidence for a quenching of the flat-spectrum jet between MJD 55217 and 55220, with the peak of the initial radio flare occurring on MJD 55221.  This radio quenching corresponds to a transition from Type C to Type A/B QPOs between MJD 55217.9 and 55218.8 \citep{Sha10}.  In the absence of VLBI monitoring close to the state transition, we cannot unequivocally relate the timing of the radio ejection event with the changes in the X-ray timing behaviour of the source.  However, it is plausible that the sequence of events was similar to that which we observed in H1743-322, namely a jet ejection immediately preceding the radio quenching, and contemporaneous with the change in the nature of the QPOs.

\subsection{The 2003 outburst of H1743-322}

As discussed in Section~\ref{sec:comparison}, the radio peak during the 2003 outburst of H1743-322 occurred on MJD 52737, several days before the steep drop in fractional rms variability \citep[occurring between MJD 52747 and 52750, according to][]{McC09}.  While \citet{McC09} do not classify the QPOs they detected, from the tabulated amplitudes, quality factors and central frequencies, the transition from Type C to Type B QPOs did not occur until at least MJD 52741.  The steep radio spectrum observed at the peak of the outburst ($\alpha=-0.62\pm0.01$ between 4.86 and 8.46\,GHz) suggests that an ejection event was indeed responsible for the peak radio emission.  However, inspecting figures 4 and 5 of \citet{McC09}, we see no evidence for a clear event in either the spectral or variability properties of the accretion flow that might have triggered this radio peak on 2003 April 8.  Furthermore, as discussed in Section~\ref{sec:comparison}, the subsequent ejection event considered by \citet{Ste12} seems to coincide with an increase in fractional rms variability and the re-appearance of canonical Type C QPOs.  Thus, the radio emission from this event does not seem to fit the pattern seen in the other outbursts discussed in this Section, as it corresponds to different X-ray timing signatures than those that appeared to be associated with jet ejection events in other sources.  However, we reiterate that the 2003 event in H1743-322 was much more complex than the 2009 outburst of this source, and the ejection event identified by \citet{Ste12} did not correspond to the first peak in the X-ray and radio light curves.  Therefore a full interpretation would also need to account for hysteresis effects (i.e.\ the time taken for the system to respond to the ongoing changes in mass accretion rate over the first few weeks of the outburst).

\subsection{Overall picture}

Jet ejection events have in the past been associated with the transition from the HIMS to SIMS, defined by a drop in the fractional rms variability and the appearance of Type B QPOs \citep{Fen09}.  Our radio monitoring of the 2009 outburst of H1743-322 shows that the peak of the integrated radio light curve is not necessarily the best indicator of the date of the radio ejection event, likely owing to the time taken for shocks to form and become optically thin.  When attempting to connect radio and X-ray behaviour that vary on timescales of days, this time delay can mask the signature of any causal connection.  This highlights the crucial role of high-cadence, high-angular resolution VLBI monitoring during state transitions in order to pin down the exact sequence of events.

In the 2009 outburst of H1743-322, we find that the jet ejection event occurred during or just before the sharp drop in the rms variability, coinciding (within the three-day uncertainty) with the disappearance of Type C QPOs and the quenching of the compact jet.  The first detection of Type B QPOs occured several days later (albeit with the caveat of the sparsely-sampled transition).  This sequence of events is approximately consistent with what was seen in the outbursts of XTE J1550-564, XTE J1859+226 and XTE J1752-223.  However, the 2002 outburst of GX339-4 suggests that the drop in rms variability is not causally connected to the radio flare, as noted by \citet{Fen09}, although the evolution away from the canonical Type C QPOs could still be associated with the radio flare.  But even this appears to be ruled out by the 2003 outburst of H1743-322.  We note that these latter two outbursts are the only ones in the sample we considered that did not appear to be preceded by any significant radio quenching.  Either the time sampling of the radio data was too sparse to detect a brief ($<1$\,d) period of quenching, or the physical properties (or, more speculatively, the launching mechanisms) of the ejection events were different in these two outbursts.

In summary, the spectral hardness at which the jets are ejected and the radio flare occurs appears to vary between outbursts (even for an individual system).  While X-ray timing studies may be better signatures of an ejection event, the exact causal sequence of events appears to vary between outbursts, as originally deduced by \citet{Fen09}.  In the subset of cases where radio quenching is observed, the ejection event seems to be connected to the point at which the Type C QPOs and band-limited noise disappear from the X-ray power density spectrum and the fractional variability begins to decrease.  While we have shown that intensive VLBI observations have the capacity to pin down the moment of ejection, dedicated coverage of new outbursts with both high angular resolution radio monitoring and X-ray timing observations is crucial to find out whether there is any common pattern to the sequence of events and hence any causal connection between the accretion and outflow phenomena.  In this light, the end of the {\it RXTE} mission implies that prior to the advent of {\it ASTROSAT} \citep{Agr06}, such studies may only be carried out with {\it XMM-Newton}.

The radio emission from the compact core jet appears to reactivate during the HIMS; it does not require the source to first reach the LHS, as appears to be necessary for the reactivation of the optical/NIR jets.  We speculate that the suppression and reactivation of the jet near the beginning and end of an outburst, respectively, occur via a gradual decrease/increase in the total jet power, over the course of several days.  If this is the case, travel-time arguments coupled with the unresolved nature of the core jets make this inconsistent with a picture in which the suppression is due to the compact core jet being switched off instantaneously at the base, with the spectral break moving to lower frequencies with time as the motion of the ejected material takes it past the surfaces of optical depth unity at each frequency.  The instantaneous suppression scenario is also ruled out by the gradual evolution of the optical/NIR emission over the course of a few days.  Since this emission comes from a region $<1$ light second downstream, suppression of the jet would instantaneously shut down the optical/NIR jet emission, which is not observed.

Reconciling our speculative scenario of gradual jet power evolution with the internal shock scenario of \citet{Fen04} would require the gradual decrease in jet power during the hard to soft state transition to be coupled with an increase in either the jet velocity or outflow flux, contradicting the naive assumption that jet power would be positively correlated with jet velocity and outflow flux.  Thus a later, separate ejection of faster-moving material might be more consistent with the observations \citep[possibly due to the ejection of the corona, as suggested by][]{Rod03,Vad03}.  This also fits in with what we know about the well-documented radio quenching prior to major radio outbursts in Cygnus X-3 \citep{Wal94}.  The jet speed in that system is thought to be high \citep[$>0.81c$][]{Mio01}, whereas the radio quenching can last for up to two months \citep{Wal94}, with the end of the quench phase being immediately followed by a major radio flare.  Thus, if the same process is responsible for the radio quenching in Cygnus X-3 and the other X-ray binaries we have studied, the onset of the quench phase cannot be directly linked to the ejection event.  A more plausible explanation might be a subsequent, higher-velocity ejection of material towards the end of the quench phase.

\section{Conclusions}

We have presented high-cadence radio and X-ray monitoring of the 2009 outburst of the black hole candidate X-ray binary system H1743-322, using data from {\it RXTE}, the VLA, VLBA and ATCA.  The X-ray behaviour of the outburst was fairly standard, following the canonical path in the hardness-intensity diagram.

Since the radio behaviour was fairly clean, likely consisting of a single bright ejection event, we were able to use our high angular resolution VLBI radio monitoring to track the proper motion of the jet components and pin down the moment of jet ejection.  The time of ejection appears to correspond, within uncertainties, to the beginning of the decrease in the fractional rms variability seen in the X-ray data, and to the disappearance of Type C quasi-periodic oscillations (i.e.\ to the transition from the hard intermediate state to the soft intermediate state).  A quenching of the radio emission occurs during or immediately after the initial ejection event, and there is a delay before the onset of the radio flare, likely due to the time taken for internal shocks to form in the outflow.  This implies that the onset of the radio flare may not always be a good diagnostic of the moment of ejection, an assumption which has frequently been made in the past and may have hampered previous attempts to couple the radio and X-ray behaviour during black hole X-ray binary outbursts.

The reactivation of the compact core jet at the end of the outburst appeared to occur during the transition from the high soft state to the hard intermediate state, and did not require the source to first reach the low hard state, as appears to be necessary for the detection of the compact jets in the optical/NIR.  This suggests that the compact jet power may evolve gradually over a period of several days during its initial suppression and subsequent reactivation near the beginning and end of an outburst, respectively.

A comparison between the outbursts of 2003 and 2009 suggests that the jet speed in H1743-322 may be variable, and possibly positively correlated with the luminosity of the outburst.  If the identification of the ejection event is correct in both outbursts, the X-ray spectral and timing signatures of the jet launching event are not constant between these two outbursts of the same source.  However, the reverse transition occurred at a very similar luminosity in both outbursts, from which we estimate a Galactic Centre distance and a relatively high-mass compact object for H1743-322.

Comparison with previous outbursts appears to validate the conclusion of \citet{Fen09} that there may be no single X-ray signature of the dramatic changes in the jet properties during a state transition.  However, more intensive radio monitoring, at high angular resolution, is required to make a conclusive statement in this regard.

\section*{Acknowledgments}

We are very grateful to the NRAO, {\it RXTE} and {\it Swift} schedulers for their flexibility and prompt responses which have made these observing campaigns feasible.  We also thank the referee, Ralph Spencer, for his constructive comments, which have helped to improve this work.  GRS acknowledges the support of an NSERC Discovery Grant.  GRS and CLS were partially supported by Chandra Grants GO0-11049X and GO0-11097X and Hubble Grants HST-GO-11679.01 and HST-GO-12012.02-A.  DMR acknowledges support from a NWO Veni Fellowship.  SC acknowledges partial funding from the European Community's Seventh Framework Programme (FP7/2007-2013) under grant agreement number ITN 215212 ``Black Hole Universe''. SMa is grateful for support from a Netherlands Organization for Scientic Research (NWO) Vidi Fellowship and from The European Community Seventh Framework Programme (FP7) under grant agreement number ITN 215212 ``Black Hole Universe''.  SMi acknowledges support by the Spanish Ministerio de Ciencia e Innovaci\'on (MICINN) under grant AYA2010-21782-C03-01, as well as financial support from MICINN and European Social Funds through a \emph{Ram\'on y Cajal} Fellowship.  The National Radio Astronomy Observatory is a facility of the National Science Foundation operated under cooperative agreement by Associated Universities, Inc.  The Australia Telescope Compact Array is part of the Australia Telescope National Facility which is funded by the Commonwealth of Australia for operation as a National Facility managed by CSIRO.  This work made use of the Swinburne University of Technology software correlator, developed as part of the Australian Major National Research Facilities Programme and operated under licence.  This research has made use of NASA's Astrophysics Data System.

\bibliographystyle{mn2e}

\label{lastpage}

\end{document}